\documentclass[12pt]{iopart}

\usepackage{amsfonts, amsbsy, amssymb, amsthm, color, url}

\usepackage[mathscr]{eucal}

\usepackage{multirow, adjustbox}

\usepackage{graphicx}

\begin{document}

\title[End-to-End Deep Learning for Interior Tomography with Low-Dose X-ray CT]{End-to-End Deep Learning \\for Interior Tomography with Low-Dose X-ray CT}

\author{Yoseob Han, Dufan Wu, Kyungsang Kim, and Quanzheng Li}

\address{Department of Radiology, Center for Advanced Medical Computing and Analysis (CAMCA), Harvard Medical School and Massachusetts General Hospital, Bostan, MA, USA}
\ead{\{yhan5, dwu6, kkim24, li.quanzheng\}@mgh.harvard.edu}
\vspace{10pt}
\begin{indented}
\item[]September 2021
\end{indented}

\begin{abstract}
\textbf{Objective:} There exist several X-ray computed tomography (CT) scanning strategies to reduce a radiation dose, such as (1) sparse-view CT, (2) low-dose CT, and (3) region-of-interest (ROI) CT (called interior tomography). To further reduce the dose, the sparse-view and/or low-dose CT settings can be applied together with interior tomography. Interior tomography has various advantages in terms of reducing the number of detectors and decreasing the X-ray radiation dose. However, a large patient or small field-of-view (FOV) detector can cause truncated projections, and then the reconstructed images suffer from severe cupping artifacts. In addition, although the low-dose CT can reduce the radiation exposure dose, analytic reconstruction algorithms produce image noise. Recently, many researchers have utilized image-domain deep learning (DL) approaches to remove each artifact and demonstrated impressive performances, and the theory of deep convolutional framelets supports the reason for the performance improvement.\\
\textbf{Approach:} In this paper, we found that the image-domain convolutional neural network (CNN) is difficult to solve coupled artifacts, based on deep convolutional framelets. \\
\textbf{Significance:} To address the coupled problem, we decouple it into two sub-problems: (i) \textbf{image domain noise reduction \textit{inside truncated projection} to solve low-dose CT problem} and (ii) \textbf{extrapolation of projection \textit{outside truncated projection} to solve the ROI CT problem}. The decoupled sub-problems are solved directly with a novel proposed end-to-end learning using dual-domain CNNs.\\
\textbf{Main results:} We demonstrate that the proposed method outperforms the conventional image-domain deep learning methods, and a projection-domain CNN shows better performance than the image-domain CNNs which are commonly used by many researchers.
\end{abstract}

%
%
%
%
%

\clearpage

\section{Introduction}
\label{sec:intro}
X-ray CT imaging provides high-quality and high-resolution images, but X-ray CT causes potential cancer risks due to radiation exposures \cite{shah2008alara}. 
Thus, many researchers have studied to reduce the radiation dose \cite{nuyts2013modelling}, where three approaches were widely used by reducing (1) projection views (sparse-view CT), (2) photon counts of X-ray source (low-dose CT), and (3) ROI (interior tomography). Unlike the low-dose CT reducing photon counts and the sparse-view CT undersampling projection views, the interior tomography retains these factors but uses small FOV detectors, which are useful for imaging of small target regions such as cardiac and dental imagings. In addition, portable C-arm CTs also use interior tomography imaging to miniaturize the hardware system. Therefore, interior tomography not only reduces the radiation exposures but also has a cost-benefit due to the small size of detectors. In addition, the sparse-view and/or low-dose settings can be applied together to interior tomography, and then the radiation exposure dose is extremely reduced compared to using interior tomography only. While interior tomography has various advantages, truncated projection data has not been correctly reconstructed using analytic CT reconstruction algorithms such as filtered backprojection (FBP) and the reconstructed image suffers from severe cupping artifacts. In addition, the reconstructed images from multiple incomplete CT measurements show additional artifacts such as streaking artifacts due to sparse-view and/or image noise caused by low-dose.

A simple method to mitigate the cupping artifacts caused by interior tomography is projection extrapolation \cite{hsieh2004algorithm}. Even though the reconstructed image using extrapolated projection data shows moderated cupping artifacts, 
Hounsfield units (HU) can be biased due to inaccurate extrapolation \cite{hsieh2004novel}. Other researchers have developed model-based iterative reconstruction (MBIR) methods with several penalty teams such as total variation (TV) \cite{yu2009compressed} and generalized L-spline \cite{ward2015interior, lee2015interior}. 
Similar to interior tomography, various MBIR methods have been investigated to address streaking \cite{pan2009commercial, bian2010evaluation, lu2011few, kim2014sparse, abbas2013effects} and image noise \cite{beister2012iterative, ramani2011splitting}. However, a drawback of MBIR-based methods is that it requires a long reconstruction time due to the computationally intensive CT operators like projector and backprojector.

\begin{figure}[t!]
  \centering
  \includegraphics[width=0.9\textwidth]{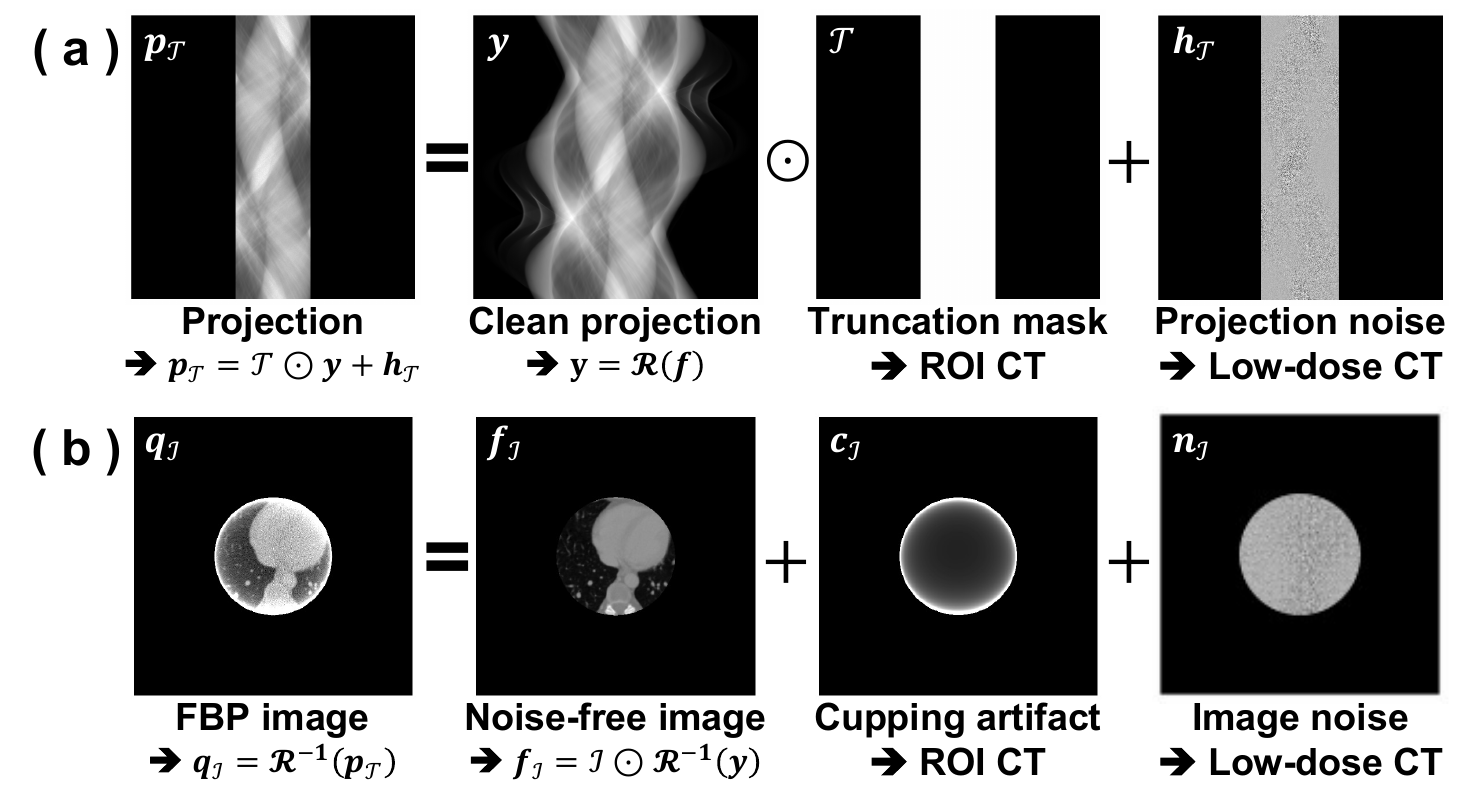}
  \caption{Low-dose ROI CT compositions of (a) measurement $p_{\mathscr{T}} = \mathscr{T} \odot y + h_{\mathscr{T}}$ in projection domain and (b) FBP image $q_{\mathscr{I}} = f_{\mathscr{I}} + c_{\mathscr{I}} + n_{\mathscr{I}}$ in image domain.}
  \label{fig:artifacts}
\end{figure}

Recently, DL algorithms have been proposed as high-performance solutions for sparse-view CT \cite{jin2017deep, lee2018deep, han2018framing}, low-dose CT \cite{wolterink2017generative, chen2017low, kang2017deep}, and interior tomography \cite{han2017deep, han2019one}. The solutions based on DL have surpassed the conventional MBIR methods \cite{yu2009compressed, ward2015interior, lee2015interior, pan2009commercial, bian2010evaluation, lu2011few, kim2014sparse, abbas2013effects, beister2012iterative, ramani2011splitting} in terms of image quality and reconstruction time. Although the DL methods have been applied to the image domain to remove each artifact and achieve excellent performance, a fundamental reason for the artifacts is incomplete measurements of a projection domain like a small number of views, photons, and small-sized detectors. In addition, if the projection $y$ is distorted by two or more collapsed factors as shown in Fig. \ref{fig:artifacts}(a), the FBP image $q_{\mathscr{I}}$ in Fig. \ref{fig:artifacts}(b) is contaminated by mixed artifacts by a cupping artifact $c_{\mathscr{I}}$ and an image noise $n_{\mathscr{I}}$.
More specifically, the cupping artifact $c_\mathscr{I}$ is described as a single global artifact but the image noise $n_\mathscr{I}$ appears as dispersion artifacts. In other words, the blended artifact ($c_\mathscr{I}$ + $n_\mathscr{I}$) has all of the opposite properties. 
Theoretically, the deep convolutional framelets \cite{ye2018deep}, which is a mathematical framework for understanding deep learning behavior, has proved that the low-rankness is a crucial factor for high performance, however, the combined opposite features are difficult to satisfy the low-rankness. Therefore, it is difficult for a single model to learn an optimal feature that contains features composed of opposite distributions.

\begin{figure}[t!]
  \centering
  \includegraphics[width=0.95\textwidth]{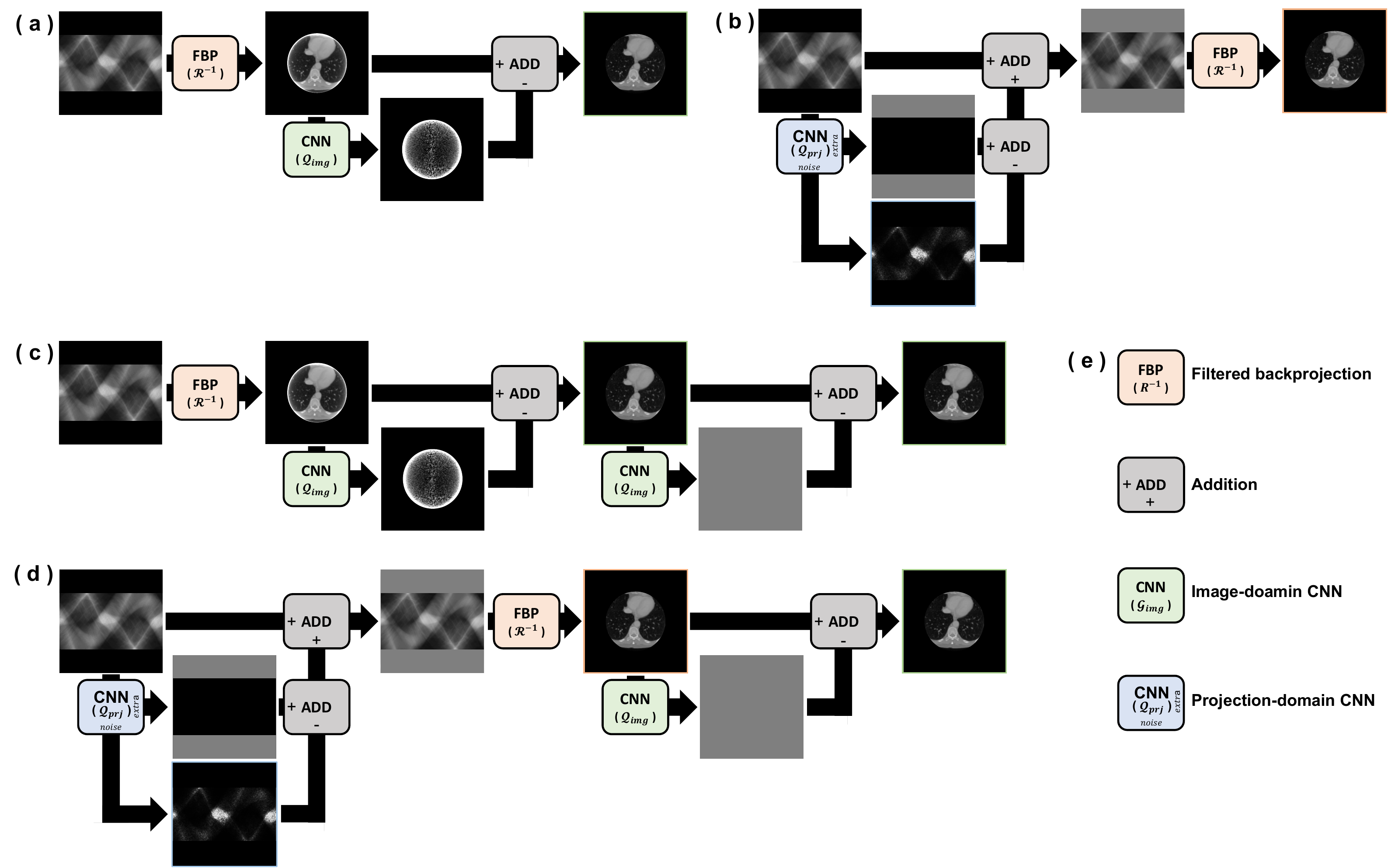}
  \caption{Various neural network architectures. (a) image-domain CNN, (b) projection-domain CNN, (c) W-Net, and (d) proposed network (called Dual-Net). (e) describes function modules used in (a-d).}
  \label{fig:architectures}
\end{figure}

In this paper, we propose a novel end-to-end deep learning method to solve a coupled low-dose ROI CT problem, simultaneously. Since a standard CNN architecture like Fig. \ref{fig:architectures}(a) is difficult to solve the coupled CT problem in an image domain due to unsatisfying a low rankness based on deep convolutional framelets \cite{ye2018deep}, a novel projection-domain CNN (see Fig. \ref{fig:architectures}(b)) is proposed to eliminate a mixture of artifacts combining the cupping artifact and the image noise by estimating two decoupled projection domain solutions: (i) \textbf{image domain noise reduction \textit{inside} a measured region $\mathscr{T}$} and (ii) \textbf{extrapolation of projection \textit{outside} the measured region $(1 - \mathscr{T})$}. After the projection-domain CNN, an image-domain CNN is applied to an image reconstructed by the projection-domain CNN. Therefore, the proposed network consists of a projection-domain CNN and an image-domain CNN, and its flowchart is illustrated in Fig. \ref{fig:architectures}(d). To compare the same architecture with the proposed network, Fig. \ref{fig:architectures}(c) (called W-Net) is used as the comparison architecture.

This paper is structured as follows. In Section \ref{sec:theory}, interior tomography and low-dose X-ray CT problems are defined, and the theory of deep convolutional framelets are briefly reviewed. Section \ref{sec:contribution} describes the limitations of conventional image-domain CNNs and how the proposed method addresses the problem when solving the low-dose ROI CT problem. Then, Section \ref{sec:method} describes the methods to implement and validate the proposed method, and experimental results are followed in Section \ref{sec:result}. Discussions and conclusions are provided in Sections \ref{sec:discussion} and \ref{sec:conclusion}.


\section{Theory}
\label{sec:theory}
\subsection{Interior tomography}
\label{sec:interior}
Here, we first describe Radon transform $\mathscr{R}$ and then extend it to the interior tomography problem using a truncated Radon transform $\mathscr{T}_{\mu}\mathscr{R}$. Let $\theta$ denotes a vector on the unit sphere $\mathbb{S} \in \mathbb{R}^2$. The set of orthogonal vectors $\theta^\perp$ is described as
\begin{eqnarray}
	\mathbf{\theta}^\perp = \{\mathbf{v} \in \mathbb{R}^2 : \mathbf{v} \cdot \mathbf{\theta} = 0\},
\end{eqnarray}
where $\cdot$ denotes an inner product. If an image is defined by $f(\mathbf{x})$ for $\mathbf{x} \in \mathbb{R}^2$, the Radon transform $\mathscr{R}$ of the image $f$ is formulated as 
\begin{eqnarray}
\label{eq:radon}
	\mathscr{R}f(\mathbf{\theta}, u) :=\int_{\mathbf{\theta}^\perp}	d\mathbf{v}~{f(\mathbf{v} + u\mathbf{\theta})},
\end{eqnarray}
where $u\in \mathbb{R}$ and $\mathbf{\theta}\in\mathbb{S}$. 
The Radon transform $\mathscr{R}$ can be reformulated as a X-ray transform $D_f$, is defined as
\begin{eqnarray}
\label{eq:line_integral}
	D_f(\mathbf{\theta}, \mathbf{a}) =\int_{0}^{\infty}{dt~f(\mathbf{a} + t \mathbf{\theta})},
\end{eqnarray}
where $\mathbf{a} \in \mathbb{R}^2$ denotes a X-ray source position. 

\begin{figure}[t!]
  \centering
  \includegraphics[width=0.3\textwidth]{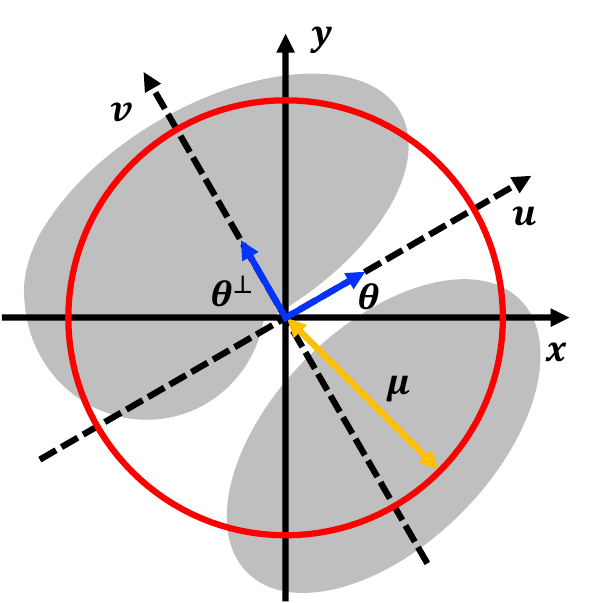}
  \caption{A CT coordinate system.}
  \label{fig:ct_geometry}
\end{figure}

If the Radon transform $\mathscr{R}f$ is restricted by $\{(\mathbf{\theta},u) : |u| < \mu \}$ 
as shown in Fig. \ref{fig:ct_geometry}, then a truncated Radon transform is defined as $\mathscr{T}_{\mu} \mathscr{R}f$, where $\mathscr{T}_{\mu}$ is truncation mask with radius $\mu$. Therefore, a interior tomography problem can be explained to find an unknown image $f(\mathbf{x})$ for $|\mathbf{x}| < \mu$ from the truncated Radon transform $\mathscr{T}_{\mu} \mathscr{R} f$.
However, the truncated Radon transform $\mathscr{T}_{\mu} \mathscr{R} f$ causes a null space $\mathscr{N}$, and the null space $\mathscr{N}$ makes the interior tomography problem a strong ill-posed problem. The null space image $\mathscr{N}$ can be illustrated like a cupping artifact $c_\mathscr{I}$ as shown in the Fig \ref{fig:artifacts}(b). Cupping artifacts appear as single global artifacts on CT images and are particularly the singularities at ROI boundaries. In particular, the artifacts are mainly related to how much of the patients are truncated and the density of the truncated tissue.

\subsection{Low-dose X-ray CT}
\label{sec:low_dose}
When a X-ray source satisfies monochromatic condition and there is no projection noise, i.e. the number of incident photons $I0$ is sufficiently large, the number of transmitted photons $I$ measured by detectors follows Poisson distribution \cite{hsieh2003computed}, is described as 
\begin{eqnarray}
\label{eq:poisson}
I \sim Poisson(I0 * \textrm{exp}^{-[D_{f}]}).
\end{eqnarray}
If the measurement $I$ is not distorted by projection noise, Eq. \ref{eq:poisson} is linearized by a negative logarithmic transform as
\begin{eqnarray}
\label{eq:linear}
y = - \ln{\left( \frac{I}{I0} \right)} = - \ln{\left( \textrm{exp}^{-[D_{f}]} \right)} = D_f = \mathscr{R}f.
\end{eqnarray}
Since a projection data $y$ satisfies Radon transform $\mathscr{R}$ of the image $f$, the image $f$ can be reconstructed analytically by applying an inverse Radon transform $\mathscr{R}^{-1}$. 

If the number of incident photons $I0$ is not enough to collect the noise-free projection data $y$, i.e. low-dose X-ray CT, the projection data $y$ is collapsed by non-stationary Gaussian noise $h$ like \cite{kang2017deep, zhu2012noise}:
\begin{eqnarray}
\label{eq:low_dose}
p \thickapprox y + h.
\end{eqnarray}
Since the collapsed projection $p$ no longer satisfies Radon transform $\mathscr{R}$ relationship of the image $f$, its reconstructed image $q$ from the collapsed projection $p$ is consisted of clean image $f$ and image noise $n$:
\begin{eqnarray}
\label{eq:fbp_low_dose}
q = \mathscr{R}^{-1}(p) \thickapprox \mathscr{R}^{-1}(y + h) = f + n.
\end{eqnarray}
As shown in Fig. \ref{fig:artifacts}(b), the image noise $n_\mathscr{I}$ appear as dispersion artifacts in all areas of the CT images, in contrast to cupping artifacts $c_\mathscr{I}$, which are a single global artifacts in interior tomography.

\begin{figure}[t!]
  \centering
  \includegraphics[width=0.8\textwidth]{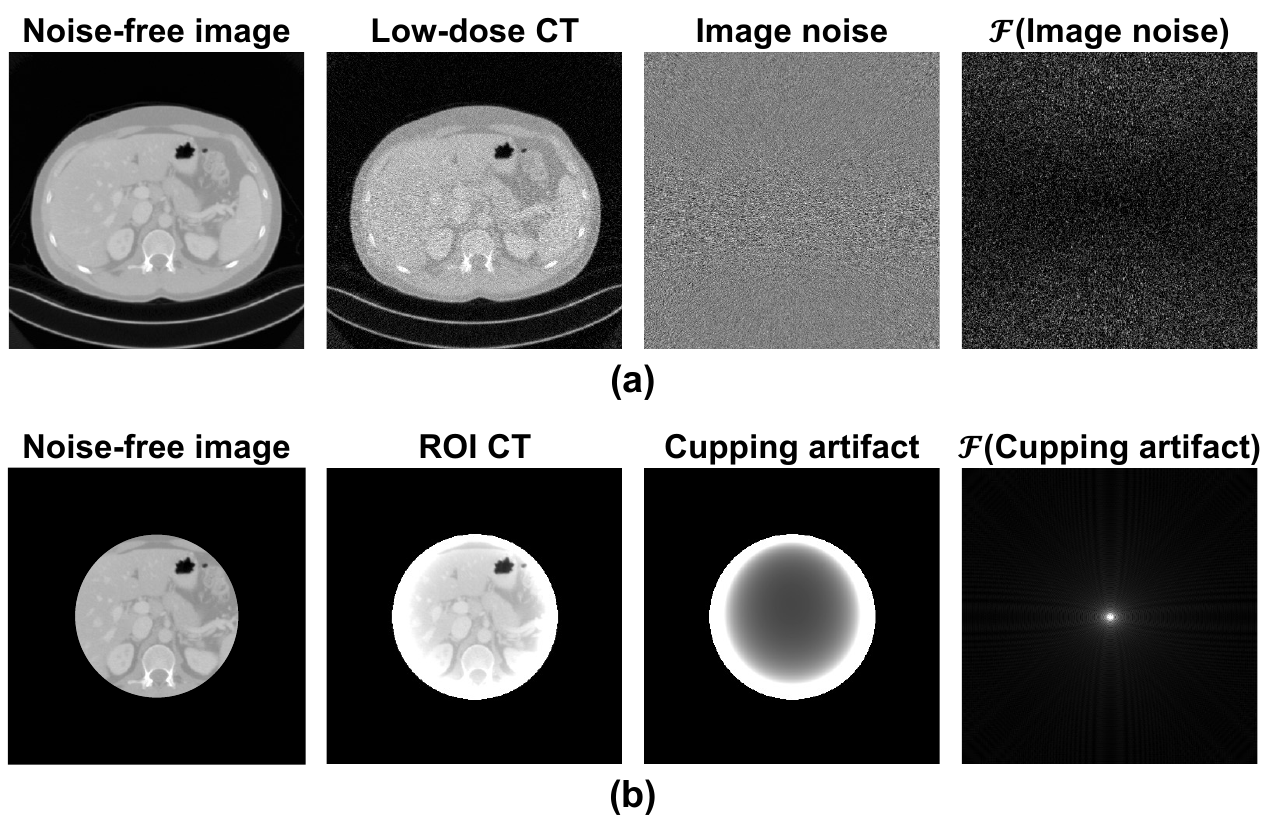}
  \caption{(a) Image noise property of low-dose CT  and (b) Cupping artifact property of ROI CT. $\mathscr{F}$ denotes 2D Fourier transform.}
  \label{fig:artifacts_spec}
\end{figure}

\subsection{Deep convolutional framelets}
Deep convolutional framelets \cite{ye2018deep, ye2019understanding} provided a mathematical link between classical signal processing and deep learning. The mathematical link is started from the Hankel matrix approaches \cite{jin2015annihilating, jin2017sparse, jin2016general, lee2016acceleration, lee2016reference}, and we start from the regression problem with low-rank Hankel structured matrix constraint defined below:
\begin{eqnarray}
\label{eq:aloha}
\arg \min_{\bar{f} \in \mathbb{R}^n}&~{|| {f - \bar{f}} ||^2} \\ \nonumber
\rm{subject~to}&~\small{\textrm{RANK}}\mathbb{H}_d(\bar{f}) = r < d,
\end{eqnarray}
where $f\in\mathbb{R}^{n}$ and $\bar{f}\in\mathbb{R}^{n}$ are a noise-free image and a denoised solution, respectively, $r$ denotes the rank of the Hankel structured matrix $\mathbb{H}_d(\bar{f}) \in \mathbb{R}^{n \times d}$, and $d$ is a matrix pencil parameter. Specifically, the rank of the Hankel structured matrix $\small{\textrm{RANK}}\mathbb{H}_d(\bar{f})$ is determined by the number of non-zero components in Fourier domain of the solution $\mathscr{F}(\bar{f})$:
\begin{eqnarray}
\label{eq:num_rank}
\small{\textrm{RANK}}\mathbb{H}_d(\bar{f}) = \small{\textrm{COUNT}}\left(\mathscr{F}(\bar{f}) \neq 0\right).
\end{eqnarray}
If there is any feasible solution $\bar{f}$, the singular value decomposition (SVD) of its Hankel structured matrix $\mathbb{H}_d(\bar{f})$ is calculated by $\textrm{SVD}(\mathbb{H}_d(\bar{f})) = U \Sigma V^T$, where $U\in\mathbb{R}^{n \times r}$ and $V\in\mathbb{R}^{d \times r}$ are the left and right singular vector bases matrices, respectively, and $\Sigma = (\sigma) \in \mathbb{R}^{r \times r}$ denotes the diagonal matrix of singular values. Here, we consider two matrices pairs $\Phi, \tilde{\Phi} \in \mathbb{R}^{n \times n}$ and $\Psi, \tilde{\Psi} \in \mathbb{R}^{d \times r}$ satisfying conditions below:
\begin{eqnarray}
\label{eq:frames}
(a)~\tilde{\Phi} \Phi^T = I_{n \times n},~~~~~(b)~\Psi \tilde{\Psi}^T = P_{R(V)},
\end{eqnarray}
where $R(V)$ denotes the range space of $V$ and $P_{R(V)}$ is a projection onto $R(V)$. Using Eq. \ref{eq:frames}, we can formulate an equality of the Hankel structured matrix $\mathbb{H}_d(\bar{f})$, is defined as:
\begin{eqnarray}
\label{eq:hankel}
\mathbb{H}_d(\bar{f}) = \tilde{\Phi} \Phi^T \mathbb{H}_d(\bar{f}) \Psi \tilde{\Psi}^T.
\end{eqnarray}
Specifically, $\Phi$ and $\tilde{\Phi}$ are called \textit{non-local bases} because they multiply by the left of $\mathbb{H}_d(\bar{f})$, then interacts with all $\bar{f}$. In common deep learning terminology, the matrices $(\Phi, \tilde{\Phi})$ correspond to user-defined general pooling $\Phi$ and unpooling $\tilde{\Phi}$. However, $\Psi$ and $\tilde{\Psi}$ are defined as \textit{local bases} since they interacts with $d$-neighborhood of the image $\bar{f}$. They are referred to as learnable kernels of the convolutional layers that can be trained to satisfy Eq. \ref{eq:frames}(b). Based on Eq. \ref{eq:hankel}, we can set a space $\mathcal{F}_r$ collecting feasible images $\bar{f}$, is described as
\begin{eqnarray}
\label{eq:set_h}
\mathcal{F}_r = \left\{ {\bar{f}\in\mathbb{R}^n \bigg| \bar{f} = \left( {\tilde{\Phi}C} \right) \circledast \nu(\tilde{\Psi}), C=\Phi^T(\bar{f}\circledast \bar{\Psi}) } \right\},
\end{eqnarray}
where $\bar{\Psi}$ and $\nu(\tilde{\Psi})$ denote encoder- and decoder-layer convolutional filters, respectively.
The previous regression problem in Eq. \ref{eq:aloha} can be reformulated using the space $\mathcal{F}_r$ given by 
\begin{eqnarray}
\label{eq:aloha_w_H}
\arg \min_{\bar{f} \in \mathcal{F}_r}{|| {f - \bar{f}} ||^2},
\end{eqnarray}
which can be represented by optimizing kernels $(\Psi, \tilde{\Psi})$ of neural network $\mathscr{Q}$ as follows
\begin{eqnarray}
\label{eq:op_kernel}
\arg \min_{(\Psi, \tilde{\Psi})}{|| {f - \mathscr{Q}(q; \Psi, \tilde{\Psi})} ||^2}.
\end{eqnarray}
The neural network $\mathscr{Q}$ can be trained with big datasets $\{ (q^{(i)}, f^{(i)}) \}_{i=1}^N$ to learn the kernels $(\Psi, \tilde{\Psi})$ representing $\small{\textrm{RANK}}\mathbb{H}_d(\bar{f}^{(i)}) \leq r_{\textrm{max}}$, where $r_{\textrm{max}}$ is the largest rank of the Hankel structured matrix $\mathbb{H}_d(\bar{f}^{(i)})$ among the datasets and $d$ is redefined by the convolutional filter length, 
is described by
\begin{eqnarray}
\label{eq:op_net}
\arg \min_{(\Psi, \tilde{\Psi})}\sum_{i=1}^{N}{|| {f^{(i)} - \mathscr{Q}(q^{(i)}; \Psi, \tilde{\Psi})} ||^2}.
\end{eqnarray}
Generally, various network architectures have been developed based on mathematical expressions or image priors, but they can be well trained with a given datasets due to their expressive power based on convolutional frames \cite{ye2018deep, ye2019understanding}. In this paper, the standard U-Net \cite{ronneberger2015u} was used as backbone architecture in order to minimize the effects of different network architectures.

\begin{figure}[t!]
  \center
  \includegraphics[width=0.8\textwidth]{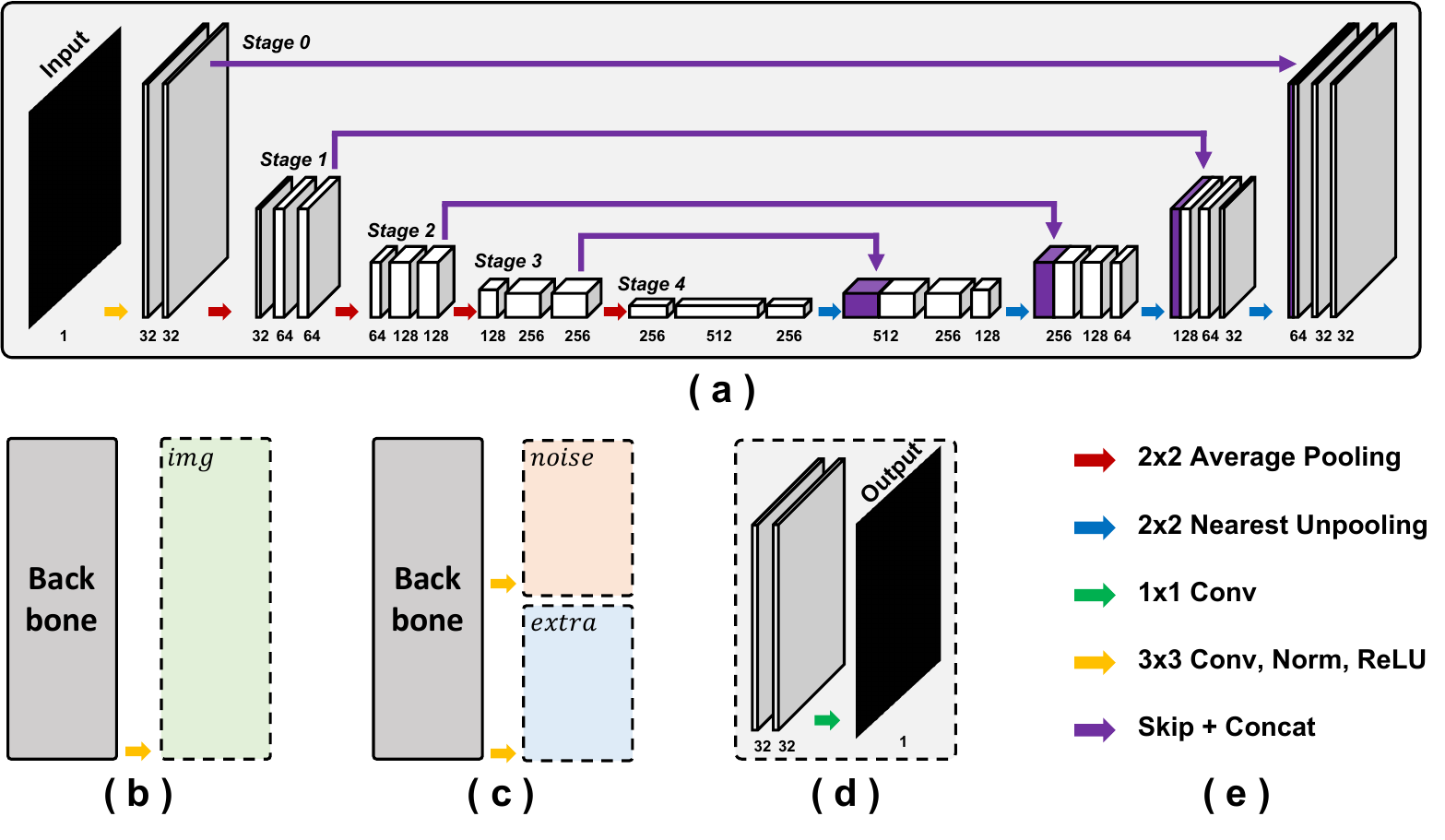}
  \caption{(a) A backbone based on the standard U-Net structure, (b) image-domain CNN $\mathscr{Q}_{img}$ consisting a backbone and a single bridge module to estimate an image, (c) projection-domain CNN $\mathscr{Q}_{prj}$ consisting the backbone and two bridge modules to estimate a projection noise inside the measured region $\mathscr{T}$ and an extrapolation map outside the measured region $(1 - \mathscr{T})$, respectively, and (d) a bridge module. (e) shows definition of layers.}
  \label{fig:backbone}
\end{figure}

\section{Materials and Methods}
\label{sec:method}
\subsection{Datasets}
Ten subject datasets from the American Association of Physicists in Medicine (AAPM) Low-Dose CT Grand Challenge \cite{mccollough2016tu} were used. Among ten subjects, nine subjects were used as training (8 subjects with 3,990 slices) and validation (1 subject with 252 slices) datasets. Another subject with 486 slices was used to test dataset. From the datasets, projection data were numerically generated using a forward projection operator with fan beam with equispaced geometry. A size of images is $512 \times 512$ and its pixel resolution is $1mm^2$. The number of views is $720$ views and a range of rotation for X-ray source is $[0^\circ,~360^\circ)$. The number of detectors is $1440$ and the detector pitch is $1mm$. The number of incident photons $(I0)$ is randomly defined as $10^{R(5, 8)}$, where $R(a, b)$ is a uniform random number generator between $a$ and $b$, and a low-dose simulation was performed according to the method of Yu \etal \cite{yu2012development}.
Truncation ratios were used as [0\%, $R(0, 58)$\%, 58\%, $R(58, 74)$\%, 74\%, $R(74, 83)$\%, 83\%], so datasets were extended seven times. Therefore, 27,930 (= 3,990 $\times$ 7) slices and 1,764 (= 252 $\times$ 7) slices are used for train and validation datasets, respectively.

For quantitative evaluation, three metrics such as the normalized mean square error (NMSE), the peak signal to noise ratio (PSNR), and the structural similarity index measure (SSIM) are used. NMSE is computed by
\begin{eqnarray}
\textrm{NMSE}(f^{*}, \bar{f}) = \frac{|| {f^{*} - \bar{f}} ||^{2}_2}{|| {f^{*}} ||^{2}_2},
\end{eqnarray}
where $f^{*}$ and $\bar{f}$ denote the ground truth and the estimated image, respectively. 
PSNR is defined as
\begin{eqnarray}
\textrm{PSNR}(f^{*}, \bar{f}) = 20 \cdot \log_{10} \left( {\frac{NM|| f^{*} ||_{\infty}}{|| f^{*} - \bar{f} ||_{2}}} \right),
\end{eqnarray}
where $N$ and $M$ are the number of pixels for row and column. SSIM is formulated as
\begin{eqnarray}
\textrm{SSIM}(f^{*}, \bar{f}) = \frac{(2\mu_{f^{*}}\mu_{\bar{f}} + c_{1})(2\sigma_{f^{*}\bar{f}} + c_{2})}{(\mu_{f^{*}}^2 + \mu_{\bar{f}}^2 + c_{1})(\sigma_{f^{*}}^2 + \sigma_{\bar{f}}^2 + c_{2})},
\end{eqnarray}
where $\mu_{f}$ and $\sigma_{f}^{2}$ are a mean and a variance for $f \in [f^{*}, \bar{f}]$, respectively, and $\sigma_{f^{*}\bar{f}}$ is a covariance of $f^{*}$ and $\bar{f}$. There are two variables to stabilize the division such as $c_{1} = (k_1L)^2$ and $c_2 = (k_2L)^2$. $L$ is a dynamic range of the pixel intensities. $k_1=0.01$ and $k_2=0.03$ are constants.

\subsection{Architectures}
Figs. \ref{fig:architectures}(a-d) illustrate an image-domain CNN (called U-Net), a projection-domain CNN, a W-Net consisting of two image-domain CNNs, and a proposed network consisting of projection-domain CNN and image-domain CNN, respectively. Details of the image-domain CNN $\mathscr{Q}_{img}$ and the projection-domain CNN $\mathscr{Q}_{prj}$ illustrated in Fig. \ref{fig:architectures} are described in Figs. \ref{fig:backbone}(b, c), respectively. In the proposed network in Fig. \ref{fig:architectures}(d), a filtered backprojection (FBP) operation is used to transform from an output of projection-domain CNN into an input of image-domain CNN.
A backbone architecture was used to standard U-Net structure \cite{ronneberger2015u} as shown in Fig. \ref{fig:backbone}(a). The CNNs are connected to additional bridge modules (see Fig. \ref{fig:backbone}(d)) after the backbone architecture. Specifically, an image-domain CNN in Fig. \ref{fig:backbone}(b) has a single bridge module for estimating the image. However, a projection-domain CNN in Fig. \ref{fig:backbone}(c) is attached to two bridge modules to estimate the projection noise inside measured area and the extrapolation map outside measured area, respectively. A basic layer module is consisted of $3\times3$ convolutional layer, batch normalization, and rectified linear unit (ReLU) as illustrated in yellow arrow of Fig. \ref{fig:backbone}(e). The basic layer module is between all blocks, but the yellow arrow has been omitted for visibility.
The W-Net and the proposed network use two networks as shown in Fig. \ref{fig:architectures}(c-d), whereas the U-Net in Fig. \ref{fig:architectures}(a) uses a single network, so the U-Net is set up to twice the size of the channels of other networks. The number of parameters of U-Net is 22,050,369, the W-Net is 11,028,034 (= 5,514,017 + 5,514,017) parameters, and the proposed network is 11,046,563 (= 5,532,546 + 5,514,017) parameters. Therefore, the number of parameters of U-Net used the doubled parameters than others.

\begin{figure}[t!]
  \centering
  \includegraphics[width=0.8\textwidth]{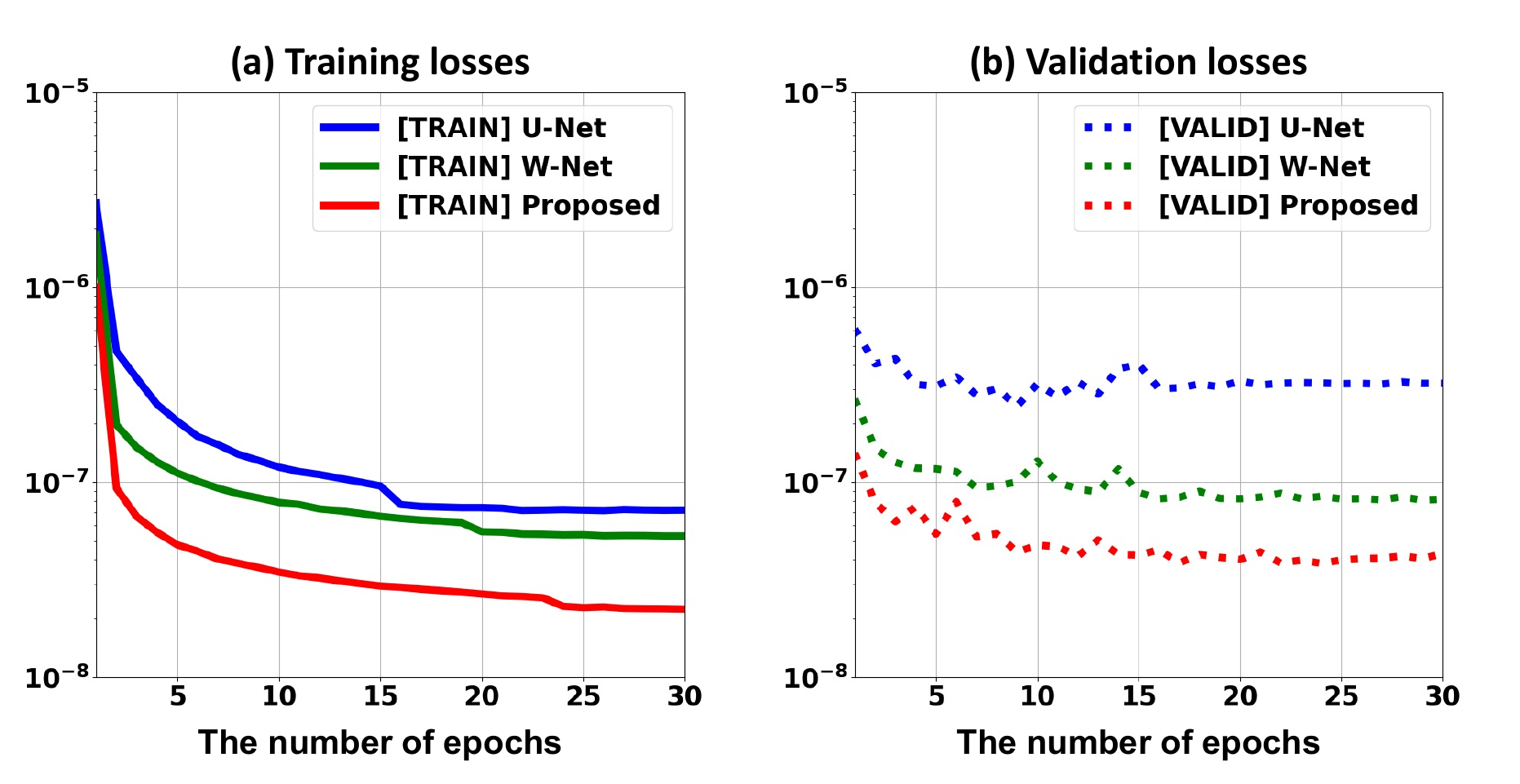}
  \caption{(a) Training losses and (b) Validation losses with respect to U-Net (blue), W-Net (green), and proposed network (red).}
  \label{fig:objective}
\end{figure}

\subsection{Training}
The networks were implemented using Pytorch. The proposed architecture was directly trained with an end-to-end learning scheme, so the FBP module was implemented as a custom layer type in Pytorch. In addition, a backward propagation of the FBP module can be sequentially conducted by the forward projection operator and the filtration. For two-times unrolled networks such as W-Net and proposed network, we blocked the gradient of second network from propagating to the first network during training phase in order to isolate the physical workspace of each network. A unit of a graphic processing unit (GPU) as NVIDIA Tesla V100 is used to train the network. Hyper parameters to train CNNs are described below. Adam optimizer was used. An initial learning rate was $10^{-4}$ and it was multiplied by 0.1 if validation loss did not decrease over 5 epochs. The number of batch sizes is 4, and the number of epochs is 30. For data augmentation, vertical flipping was applied. The above hyper parameters were applied equally to all networks as shown in Fig. \ref{fig:architectures}(a,c,d) when trained. 

\section{Main contributions}
\label{sec:contribution}

\subsection{Image-domain CNN for coupled artifact and its Limitation}
\label{sec:dif_sol_img}
The FBP image $q^{(i)}$ can be distorted by artifacts related to CT systems given by
\begin{eqnarray}
q^{(i)} = f^{(i)} + g^{(i)}.
\end{eqnarray}
where $g^{(i)} \in [c^{(i)}_\mathscr{I}, n^{(i)}]$ denotes artifacts according to incomplete CT systems like interior tomography and low-dose CT, $*_{\mathscr{I}}$ is any images applied to ROI mask ${\mathscr{I}}$, and $c^{(i)}_\mathscr{I}$ and $n^{(i)}$ are cupping artifacts and image noise, respectively. 
More specifically, the cupping artifact $c^{(i)}_\mathscr{I}$ is described as a single global artifact on CT images and the image noise $n^{(i)}$ appears as dispersion artifacts in all area of the CT images.
The artifacts have been individually conquered by various deep learning methods \cite{wolterink2017generative, chen2017low, kang2017deep, han2017deep, han2019one}, and most of them follow a residual learning, instead of Eq. \ref{eq:op_net}, given by
\begin{eqnarray}
\label{eq:op_residual}
\arg \min_{(\Psi, \tilde{\Psi})}\sum_{i=1}^{N}{|| {g^{(i)} - \mathscr{Q}(q^{(i)}; \Psi, \tilde{\Psi})} ||^2}.
\end{eqnarray}
The residual learning can be interpreted to optimize the kernel $\bar{\Psi}$ which almost annihilate the noise-free image $f^{(i)}$ like
\begin{eqnarray}
\label{eq:annihilate}
f^{(i)} \circledast \bar{\Psi} \simeq 0,
\end{eqnarray}
which is applied to Eq. \ref{eq:set_h}, then the artifacts $g^{(i)}$ can be reconstructed based on the theory of deep convolutional framelets, is represented by
\begin{eqnarray}
\label{eq:recon_artifact}
&&\left( {\tilde{\Phi} \left[ \Phi^T \left( [f^{(i)} + g^{(i)}] \circledast \bar{\Psi} \right) \right] } \right) \circledast \nu(\tilde{\Psi})\\ \nonumber
&\simeq&\left( {\tilde{\Phi} \left[ \Phi^T \left( g^{(i)} \circledast \bar{\Psi} \right) \right] } \right) \circledast \nu(\tilde{\Psi}) \\ \nonumber
&=&g^{(i)}.
\end{eqnarray}
The proof shows that the neural network $\mathscr{Q}$ with kernels $(\Psi, \tilde{\Psi})$ is well trained to represent individual artifact $g^{(i)}$, which is the cupping artifact with a dominant low-frequency Fourier support $\mathscr{F}_{\textrm{low}} (c^{(i)}_\mathscr{I})$ (see Fig. \ref{fig:artifacts_spec}(a)) or the image noise with a dominant high-frequency Fourier support $\mathscr{F}_{\textrm{high}} (n^{(i)})$ (see Fig. \ref{fig:artifacts_spec}(b)), satisfying $\small{\textrm{RANK}} \mathbb{H}_d(\bar{g}^{(i)}) \leq r_{\textrm{max}}$ where $\bar{g}^{(i)}$ is any feasible noise. 

Here, we consider that the neural network $\mathscr{Q}$ is fixed and recovers the coupled artifact $k^{(i)} = c^{(i)}_\mathscr{I} + n^{(i)}$ instead of the individual artifact $g^{(i)}$, and the cupping artifact $c^{(i)}_\mathscr{I}$ and the image noise $n^{(i)}$ with opposite Fourier supports are coupled together as shown in Fig. \ref{fig:artifacts}(b). 
Then, the kernels $(\Psi, \tilde{\Psi})$ do not satisfy with Eq. \ref{eq:frames}(b) since $\small{\textrm{RANK}}\mathbb{H}_d(\bar{k})$ 
is greater than the upper bound of the rank $r_{\textrm{max}}$, i.e.
\begin{eqnarray}
\small{\textrm{RANK}}\mathbb{H}_d(\bar{n}) \leq r_{\textrm{max}} < \small{\textrm{RANK}}\mathbb{H}_d(\bar{k}) \leq r_{\textrm{max}}^* \simeq d, \\ \nonumber
\textrm{(resp. } \small{\textrm{RANK}}\mathbb{H}_d(\bar{c}_{\mathscr{I}}) \leq r_{\textrm{max}} < \small{\textrm{RANK}}\mathbb{H}_d(\bar{k})
\leq r_{\textrm{max}}^* \simeq d \textrm{ )}
\end{eqnarray}
where $\bar{k}$ is any feasible coupled noise and $r_{\textrm{max}}^*$ denotes the upper bound of the rank of Hankel structured matrix constructed by the coupled noise $\bar{k}$. 
Since the Fourier support of the coupled noise $k^{(i)} \in \mathbb{R}^n$ fills most of the Fourier domain, its rank of the Hankel structured matrix is close to the length of the signal based on Eq. \ref{eq:num_rank}. To match the low-rank condition, the convolutional filter length $d$ must exceed the length of the signal, but it is not efficient to build a network architecture that satisfies the filter length $d$ equal to the signal length.
Therefore, the neural network $\mathscr{Q}$ no longer satisfies the deep convolutional framelets and fails to reconstruct a clean image $\bar{f}_{\mathscr{I}}=q_{\mathscr{I}}-\bar{k}$ from a multi-noisy image $q_{\mathscr{I}} = f_{\mathscr{I}} + k$.

\begin{table*}[t!]
\caption{\bf\scriptsize Objective functions in terms of various network architectures.}
\vspace*{-0.5cm}
\label{tbl:losses}
\begin{center}
\begin{adjustbox}{width=0.95\textwidth}
\begin{tabular}[t]{clclcl}

\hline \\

\multicolumn{1}{c}{\multirow{4}{*}{\rotatebox[origin=c]{0}{(a)}}} &
\multicolumn{1}{c}{\multirow{4}{*}{\rotatebox[origin=c]{0}{\textit{Symbols}}}} & 
$\odot$ & Hadamard product & $\mathscr{R}^{-1}$ & inverse Radon transform \\ \\
& & 
$\mathscr{I}$ & ROI mask in image domain & $\mathscr{T}$ & truncation mask in projection domain \\ \\
& & 
$f$ & noise-free image & $q$ & FBP image \\ \\
& &
$p$ & collapsed projection & $h$ & projection noise \\ \\ 
& & 
$n$ & image noise  & $c$ & cupping artifact \\ \\
& & 
$\Psi$ & encoder in backbone network & $\tilde{\Psi}$ & decoder in backbone network \\ \\
& & 
$\tilde{\Psi}_{*}$ & bridge modules $* \in [noise, extra]$ connected to decoder & $\mathscr{Q}_{img}$ & image-domain CNN \\ \\
& &
$\mathscr{Q}_{prj}^{*}$ & projection-domain CNN with bridge module $* \in [noise, extra]$ & & \\ \\ \hline \\

\multicolumn{1}{c}{\multirow{1}{*}{\rotatebox[origin=c]{0}{(b)}}} & 
\multicolumn{1}{c}{\multirow{1}{*}{\rotatebox[origin=c]{0}{$\mathscr{L}_{\textrm{U-Net}}^{img}$}}} & 
\multicolumn{4}{l}{$\min_{(\Psi, \tilde{\Psi}, \tilde{\Psi}_{img})}{\sum_{i=1}^{N}{\bigg|\bigg| {\mathscr{I}^{(i)} \odot \left(q^{(i)} - f^{(i)}\right) - \mathscr{I}^{(i)} \odot \mathscr{Q}_{img} \left(q^{(i)}; \Psi, \left[\tilde{\Psi}, \tilde{\Psi}_{img}\right]\right)}\bigg|\bigg|^2}}$.} \\ \\ \hline \\

\multicolumn{1}{c}{\multirow{2}{*}{\rotatebox[origin=c]{0}{(c)}}} & 
\multicolumn{1}{c}{\multirow{2}{*}{\rotatebox[origin=c]{0}{$\mathscr{L}_{\textrm{U-Net}}^{prj}$}}} & 
\multicolumn{4}{l}{$\min_{(\Psi, \tilde{\Psi}, \tilde{\Psi}_{noise})}{\sum_{i=1}^{N}{\bigg|\bigg| {\mathscr{T}^{(i)} \odot \left( p^{(i)} - y^{(i)} \right) - \mathscr{T}^{(i)} \odot \mathscr{Q}_{prj}^{noise} \left(p^{(i)}; \Psi, \left[\tilde{\Psi}, \tilde{\Psi}_{noise}\right]\right)}\bigg|\bigg|^2}}$}\\
& & 
\multicolumn{4}{l}{$+ \min_{(\Psi, \tilde{\Psi}, \tilde{\Psi}_{extra})}{\sum_{i=1}^{N}{\bigg|\bigg| {\mathscr{I}^{(i)} \odot f^{(i)} - \mathscr{I}^{(i)} \odot \mathscr{R}^{-1}  \left( \mathscr{T}^{(i)} \odot \left( p^{i} - \bar{h}^{(i)}\right) + \left(1 - \mathscr{T}^{(i)} \right) \odot \mathscr{Q}_{prj}^{extra}\left(p^{(i)}; \Psi, \left[\tilde{\Psi}, \tilde{\Psi}_{extra}\right]\right)\right) \bigg|\bigg|^2}}}$.} \\ \\ \hline \\

\multicolumn{1}{c}{\multirow{3}{*}{\rotatebox[origin=c]{0}{(d)}}} & 
\multicolumn{1}{c}{\multirow{3}{*}{\rotatebox[origin=c]{0}{$\mathscr{L}_{\textrm{W-Net}}$}}} & 
\multicolumn{4}{l}{$\min_{(\Psi^1, \tilde{\Psi}^1, \tilde{\Psi}_{img}^1)}{\sum_{i=1}^{N}{\bigg|\bigg| {\mathscr{I}^{(i)} \odot \left( q^{(i)} - f^{(i)} \right) - \mathscr{I}^{(i)} \odot \mathscr{Q}_{img^1} \left(q^{(i)}; \Psi^1, \left[\tilde{\Psi}^1, \tilde{\Psi}_{img}^1\right]\right)}\bigg|\bigg|^2}}$}\\
& & 
\multicolumn{4}{l}{$+ \min_{(\Psi^2, \tilde{\Psi}^2, \tilde{\Psi}_{img}^2)}{\sum_{i=1}^{N}{\bigg|\bigg| {\mathscr{I}^{(i)} \odot \left( \bar{q}^{(i)} - f^{(i)} \right) - \mathscr{I}^{(i)} \odot \mathscr{Q}_{img^2} \left(\bar{q}^{(i)}; \Psi^2, \left[\tilde{\Psi}^2, \tilde{\Psi}_{img}^2\right]\right)}\bigg|\bigg|^2}}$,}\\ \\
& & 
\multicolumn{4}{l}{where $\bar{q}$ denotes output of the first image-domain CNN $\mathscr{Q}_{img^1} \left(q^{(i)}; \Psi^1, \left[\tilde{\Psi}^1, \tilde{\Psi}_{img}^1\right]\right)$.} \\ \\ \hline \\

\multicolumn{1}{c}{\multirow{5}{*}{\rotatebox[origin=c]{0}{(e)}}} &
\multicolumn{1}{c}{\multirow{5}{*}{\rotatebox[origin=c]{0}{$\mathscr{L}_{\textrm{Dual-Net}}$}}} & 
\multicolumn{4}{l}{$\min_{(\Psi^1, \tilde{\Psi}^1, \tilde{\Psi}_{noise}^1)}{\sum_{i=1}^{N}{\bigg|\bigg| {\mathscr{T}^{(i)} \odot \left( p^{(i)} - y^{(i)} \right) - \mathscr{T}^{(i)} \odot \mathscr{Q}_{prj^1}^{noise} \left(p^{(i)}; \Psi^1, \left[\tilde{\Psi}^1, \tilde{\Psi}_{noise}^1\right]\right)}\bigg|\bigg|^2}}$}\\
& & 
\multicolumn{4}{l}{$+ \min_{(\Psi^1, \tilde{\Psi}^1, \tilde{\Psi}_{extra}^1)}{\sum_{i=1}^{N}{\bigg|\bigg| {\mathscr{I}^{(i)} \odot f^{(i)} - \mathscr{I}^{(i)} \odot \mathscr{R}^{-1} \left( \mathscr{T}^{(i)} \odot \left( p^{i} - \bar{h}^{(i)}\right) + \left(1 - \mathscr{T}^{(i)} \right) \odot \mathscr{Q}_{prj^1}^{extra}\left(p^{(i)}; \Psi^1, \left[\tilde{\Psi}^1, \tilde{\Psi}_{extra}^1\right]\right)\right) \bigg|\bigg|^2}}}$}\\
& & 
\multicolumn{4}{l}{$+\min_{(\Psi^2, \tilde{\Psi}^2, \tilde{\Psi}_{img}^2)}{\sum_{i=1}^{N}{\bigg|\bigg| {\mathscr{I}^{(i)} \odot \left( \bar{q}^{(i)} - f^{(i)} \right) - \mathscr{I}^{(i)} \odot \mathscr{Q}_{img^2} \left(\bar{q}^{(i)}; \Psi^2, \left[\tilde{\Psi}^2, \tilde{\Psi}_{img}^2\right]\right)}\bigg|\bigg|^2}}$,} \\ \\
& & 
\multicolumn{4}{l}{where $\bar{h} = \mathscr{Q}_{prj^1}^{noise} \left(p^{(i)}; \Psi^1, \left[\tilde{\Psi}^1, \tilde{\Psi}_{noise}^1\right]\right)$ denotes estimated projection noise from $\mathscr{Q}_{prj}^{noise}$ and} \\
& & 
\multicolumn{4}{l}{ $\bar{q} = \mathscr{R}^{-1} \left( \mathscr{T}^{(i)} \odot \left( p^{i} - \bar{h}^{(i)}\right) + \left(1 - \mathscr{T}^{(i)} \right) \odot \mathscr{Q}_{prj^1}^{extra}\left(p^{(i)}; \Psi^1, \left[\tilde{\Psi}^1, \tilde{\Psi}_{extra}^1\right]\right)\right)$ denotes estimated FBP image from $\mathscr{Q}_{prj}^{extra}$.} \\ 

\\ \hline \\

\end{tabular}
\end{adjustbox}
\end{center}
\end{table*}

\subsection{End-to-end deep learning using dual-domain CNN}
Then, is there an algorithm that satisfies deep convolutional framelets even though the image $q^{(i)}$ is contaminated by artifacts with both opposite properties? An easy way to get a solution to the question is to separate $T(n^{(i)}_{\mathscr{I}}) \in \mathscr{S}_{n}$ and $T(c^{(i)}_{\mathscr{I}}) \in \mathscr{S}_{c}$ using any transform function $T$ for the coupled artifacts $n^{(i)}_{\mathscr{I}} + c^{(i)}_{\mathscr{I}}$ in image domain, where $\mathscr{S}_{n}$ and $\mathscr{S}_{c}$ are domains transformed from the image noise $n^{(i)}$ and the cupping artifact $c^{(i)}$ using the transform function $T$, respectively, and are domains that do not overlap each other. In projection domain, the above condition is exactly achieved, the image noise $n^{(i)}_\mathscr{I}$ is transformed to 
the projection noise $h_\mathscr{T}$ inside a truncation mask $\mathscr{T}$, and the cupping artifact $c^{(i)}_\mathscr{I}$ is described as a null space outside the truncation mask $(1 - \mathscr{T})$. In other words, the low-dose CT problem is modified from removing the image noise $n^{(i)}$ in image domain (see LHS of Eq. \ref{eq:rnk_low_dose_prj}) to eliminating 
the projection noise $h^{(i)}$ in projection domain (see RHS of Eq. \ref{eq:rnk_low_dose_prj}) and can be represented by
\begin{eqnarray}
\label{eq:rnk_low_dose_prj}
\arg \min_{\bar{n}_\mathscr{I} \in \mathcal{N}_r} ||n_\mathscr{I} - \bar{n}_\mathscr{I}||^2_{\mathscr{I} = \textbf{1}} \leadsto \arg \min_{\bar{h}_\mathscr{I} \in \mathcal{H}_r} ||h_\mathscr{T} - \bar{h}_\mathscr{T}||^2_{\mathscr{T} = \textbf{1}}.
\end{eqnarray}
The interior tomography can be also represented from eliminating the cupping artifact $c^{(i)}_\mathscr{I}$ of image domain (see LHS of Eq. \ref{eq:rnk_roi_prj}) to estimating the feasible function $\bar{z}$ of projection domain outside truncated region $(1 - \mathscr{T})$ (see RHS of Eq. \ref{eq:rnk_roi_prj}), given by
\begin{eqnarray}
\label{eq:rnk_roi_prj}
\arg \min_{\bar{c}_\mathscr{I} \in \mathcal{C}_r} ||c_\mathscr{I} - \bar{c}_\mathscr{I}||^2  \leadsto \arg \min_{\bar{z}_{(1 - \mathscr{T})} \in \mathcal{Z}_r} ||f_\mathscr{I} - \mathscr{R}^{-1}_\mathscr{I}\left(y_\mathscr{T} + \bar{z}_{(1 - \mathscr{T})}\right)||^2.
\end{eqnarray}
In addition, its coupled problem can be represented by
\begin{eqnarray}
\label{eq:rnk_both_prj}
&&\arg \min_{\bar{g}_\mathscr{I} \in \mathcal{G}_r} ||(n_\mathscr{I} + c_\mathscr{I}) - \bar{g}_\mathscr{I}||^2 \\ \nonumber
&\leadsto& \arg \min_{\bar{h}_\mathscr{I} \in \mathcal{H}_r} ||h_\mathscr{T} - \bar{h}_\mathscr{T}||^2 + 
\arg \min_{\bar{z}_{(1 - \mathscr{T})} \in \mathcal{Z}_r} ||f_\mathscr{I} - \mathscr{R}^{-1}_\mathscr{I}\left(y_\mathscr{T} + \bar{z}_{(1 - \mathscr{T})}\right)||^2.
\end{eqnarray}
Most importantly, while both the cupping artifact $c_\mathscr{I}$ and the image noise $n_\mathscr{I}$ exist inside ROI in image domain, each artifact is separated inside and outside the measured region $\mathscr{T}$ in projection domain. Because of this, the capacity of the neural network $\mathscr{Q}$ is fixed, but it is possible to apply low-rank constraints to individual regions while satisfying the theory of the deep convolutional framelets.

To enforce the condition represented by the right-hand side of Eq. \ref{eq:rnk_both_prj}, we propose a projection-domain CNN $\mathscr{Q}_{prj}$ in Fig. \ref{fig:architectures}(b) and its objective function is defined in Table. \ref{tbl:losses}(c). The objective function consists of two terms, and the first term leads to an estimation of projection noise using $\mathscr{Q}_{prj}^{noise}$ inside the projection data but the second term induces extrapolation using $\mathscr{Q}_{prj}^{extra}$ outside the projection data to eliminate the cupping artifacts in image domain. In addition, an unrolled network scheme is applied after a single network such as the image-domain CNN $\mathscr{Q}_{img}$ and the projection-domain CNN $\mathscr{Q}_{prj}$ in order to improve performance and its stability \cite{jiao2021dual}. In the paper, we used an image-domain CNN $\mathscr{Q}_{img^{2}}$ as a second network in an unrolled scheme because a count-domain CNN paper \cite{eo2018kiki} has shown that the CNN unfolded in the measurement domain is not superior to the CNN unfolded in both the image domain and the count domain.
Specifically, the proposed network in Fig. \ref{fig:architectures}(d) is unrolled in different domains, and the first is the projection domain and the second is the image domain. Details of the objective function of the proposed network are described in Table. \ref{tbl:losses}(e). Similar to the proposed network, a W-Net in Fig. \ref{fig:architectures}(c) is the two-times unrolled version of the image-domain CNN $\mathscr{Q}_{img}$, and its objective function is formulated in Table. \ref{tbl:losses}(d).

\begin{figure*}[t!]
  \centering
  \includegraphics[width=0.9\textwidth]{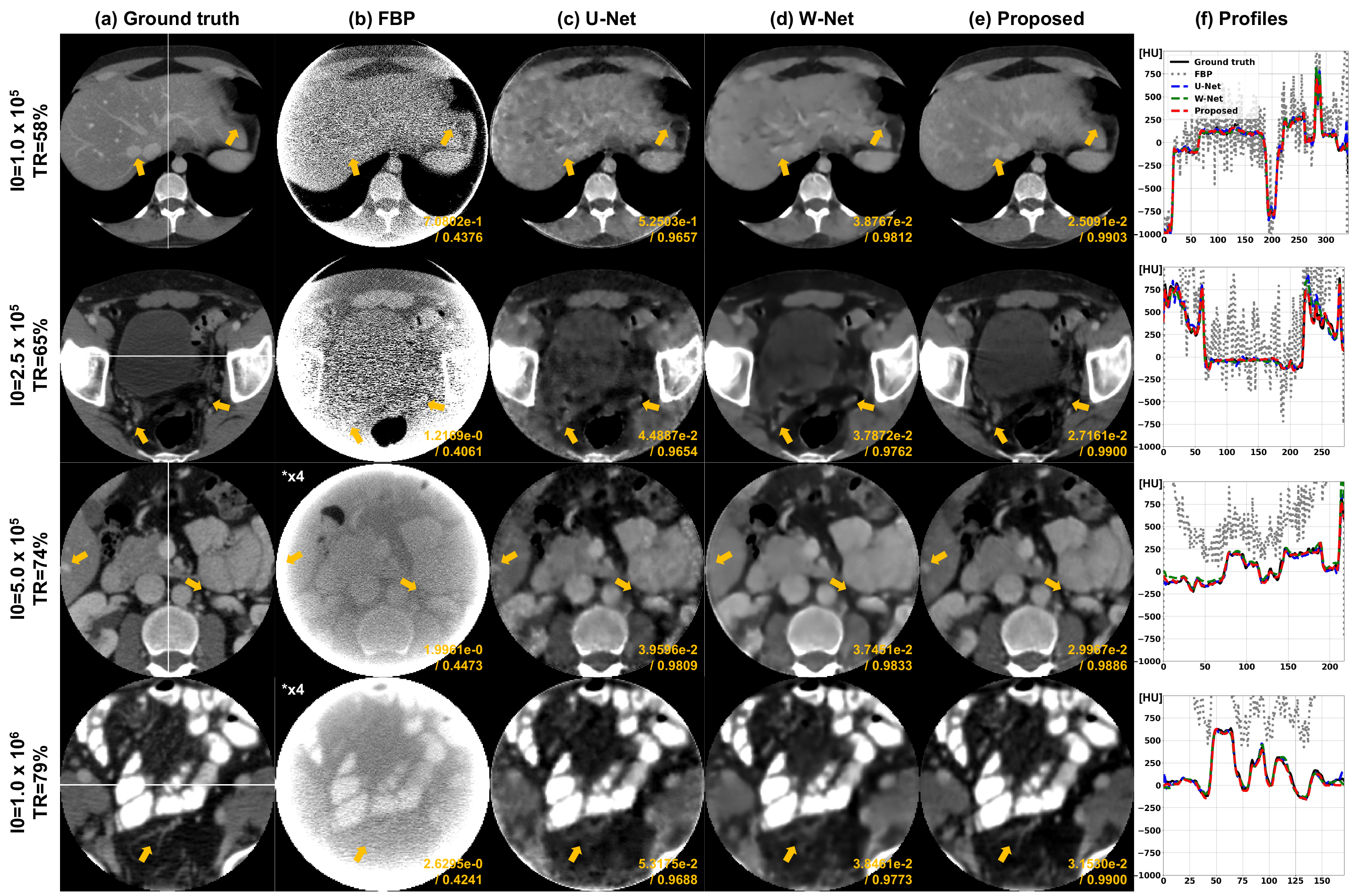}
  \caption{(a) Ground truth and reconstructed images by (b) FBP, (c) U-Net, (d) W-Net, and (e) proposed method. (f) Profiles along the white line on the results. From top to bottom, the number of photons gradually decreases, but the truncated ratio increases. The intensity range was set to $(-150, 400)$[HU]. $^{*}$x4 denotes that a window scale is magnified four times. NMSE / SSIM values are written at the corner.}
  \label{fig:result}
\end{figure*}

\section{Results}
\label{sec:result}
Fig. \ref{fig:objective} shows the objective functions with respect to the trained networks such as U-Net, W-Net, and proposed network. Even though the U-Net has double-sized parameters than others, the objective curve sits on top of other curves. The W-Net and the proposed network have a similar number of parameters and their flowcharts except for the data domain applied to the first network, but the objective curve of the proposed network shows the lowest curve than the W-Net. Therefore, the projection-domain CNN is useful for solving a coupled low-dose ROI CT problem. Table. \ref{tbl:result} shows the average NMSE and SSIM values of CNNs when applied to various low-dose ROI CTs. 
The proposed method produces the best NMSE and SSIM values than other methods. Next, the W-Net followed the performance of the proposed method, and U-Net had the lowest improvement over the others. However, when a label with an infinite number of incident photons and truncated ratio of 0\% is used as an input, the W-Net has been showed to perform worse than U-Net due to overestimation as shown in Table. \ref{tbl:result}. Interestingly, the U-Net has twice the parameters of the others but has the lowest improvement. The trade-offs between the number of parameters and unrolled network architecture will be discussed later in Section \ref{sec:unrolled}.

\begin{table*}[t!]
\caption{\bf\scriptsize Quantitative comparison with respect to various ROIs and \# of photons.}
\vspace*{-0.3cm}
\label{tbl:result}
\begin{center}
\begin{adjustbox}{width=0.65\textwidth}
\begin{tabular}{cclcccccc}
\hline\hline

\multicolumn{3}{c}{\multirow{2}{*}{NMSE $(\times 10^{-2})$}}		& \multicolumn{6}{c}{Ratios of truncated detectors} \\
\multicolumn{3}{c}{}												& $0\%$ 			& $44\%$ 			& $58\%$ 			& $65\%$			& $74\%$			& $79\%$			\\ \hline 
\multirow{24}{*}{\rotatebox[origin=c]{90}{The number of incident photons}}
& \multicolumn{1}{c}{\multirow{4}{*}{\rotatebox[origin=c]{90}{$\infty$}}}
														& FBP   	& -					& 24.025 			& 74.370			& 117.79			& 193.58			& 270.48			\\
& \multicolumn{1}{c}{}									& U-Net		& 0.9774			& 1.6771			& 2.1493			& 3.0715			& 4.1914			& 7.8876			\\
& \multicolumn{1}{c}{}									& W-Net		& 1.0173			& 1.5146			& 1.7644			& 2.0412			& 2.4303			& 3.7594			\\
& \multicolumn{1}{c}{}									& Proposed	& \textbf{0.7106}	& \textbf{0.9121}	& \textbf{1.1092}	& \textbf{1.2532}	& \textbf{2.0675}	& \textbf{2.9322}	\\ \cline{3-9} 

& \multicolumn{1}{c}{\multirow{4}{*}{\rotatebox[origin=c]{90}{$1.0 \times 10^7$}}}	
														& FBP   	& 3.5022			& 24.260 			& 74.430			& 117.83			& 193.61			& 270.50			\\
& \multicolumn{1}{c}{}									& U-Net		& 1.5299			& 1.8806			& 2.3331			& 3.2274			& 4.2828			& 7.9061			\\
& \multicolumn{1}{c}{}									& W-Net		& 1.5006			& 1.6979			& 1.9421			& 2.1864			& 2.5682			& 3.8554			\\
& \multicolumn{1}{c}{}									& Proposed	& \textbf{1.0395}	& \textbf{1.1016}	& \textbf{1.2581}	& \textbf{1.3891}	& \textbf{2.1690}	& \textbf{3.0162}	\\ \cline{3-9} 

& \multicolumn{1}{c}{\multirow{4}{*}{\rotatebox[origin=c]{90}{$1.0 \times 10^6$}}}	
														& FBP   	& 1.1332			& 26.379 			& 75.026			& 118.23			& 193.89			& 270.72			\\
& \multicolumn{1}{c}{}									& U-Net		& 2.4703			& 2.5721			& 2.9801			& 3.7945			& 4.6882			& 8.1475			\\
& \multicolumn{1}{c}{}									& W-Net		& 2.2903			& 2.2889			& 2.4881			& 2.6984			& 3.0405			& 4.2085			\\
& \multicolumn{1}{c}{}									& Proposed	& \textbf{1.5485}	& \textbf{1.5067}	& \textbf{1.6520}	& \textbf{1.7864}	& \textbf{2.4887}	& \textbf{3.2861}	\\ \cline{3-9} 

& \multicolumn{1}{c}{\multirow{4}{*}{\rotatebox[origin=c]{90}{$5.0 \times 10^5$}}}	
														& FBP   	& 15.950			& 28.457 			& 75.680			& 118.68			& 194.20			& 270.97			\\
& \multicolumn{1}{c}{}									& U-Net		& 2.8875			& 2.9551			& 3.3360			& 4.1162			& 4.9620			& 8.2956			\\
& \multicolumn{1}{c}{}									& W-Net		& 2.6322			& 2.6075			& 2.7797			& 2.9884			& 3.3216			& 4.4191			\\
& \multicolumn{1}{c}{}									& Proposed	& \textbf{1.7481}	& \textbf{1.6886}	& \textbf{1.8467}	& \textbf{1.9885}	& \textbf{2.6662}	& \textbf{3.4337}	\\ \cline{3-9} 

& \multicolumn{1}{c}{\multirow{4}{*}{\rotatebox[origin=c]{90}{$2.5 \times 10^5$}}}	
														& FBP   	& 22.427			& 32.085 			& 76.961			& 119.56			& 194.81			& 271.47			\\
& \multicolumn{1}{c}{}									& U-Net		& 3.4469			& 3.4788			& 3.8156			& 4.5546			& 5.4141			& 8.6357			\\
& \multicolumn{1}{c}{}									& W-Net		& 3.0635			& 3.0336			& 3.1664			& 3.3856			& 3.6960			& 4.7537			\\
& \multicolumn{1}{c}{}									& Proposed	& \textbf{1.9800}	& \textbf{1.9049}	& \textbf{2.0818}	& \textbf{2.2311}	& \textbf{2.8875}	& \textbf{3.6211}	\\ \cline{3-9} 

& \multicolumn{1}{c}{\multirow{4}{*}{\rotatebox[origin=c]{90}{$1.0 \times 10^5$}}}	
														& FBP   	& 34.959			& 40.694 			& 80.618			& 122.15			& 196.66			& 272.94			\\
& \multicolumn{1}{c}{}									& U-Net		& 4.6303			& 4.5646			& 4.7933			& 5.5142			& 6.5089			& 9.6548			\\
& \multicolumn{1}{c}{}									& W-Net		& 3.8775			& 3.8302			& 3.9137			& 4.1427			& 4.4446			& 5.5535			\\
& \multicolumn{1}{c}{}									& Proposed	& \textbf{2.3678}	& \textbf{2.2633}	& \textbf{2.4671}	& \textbf{2.6180}	& \textbf{3.2606}	& \textbf{3.9583}	\\ \cline{3-9} 
\hline \hline


\multicolumn{3}{c}{\multirow{2}{*}{SSIM}}							& \multicolumn{6}{c}{Ratios of truncated detectors} \\
\multicolumn{3}{c}{}												& $0\%$ 			& $44\%$ 			& $58\%$ 			& $65\%$			& $74\%$			& $79\%$			\\ \hline 
\multirow{24}{*}{\rotatebox[origin=c]{90}{The number of incident photons}}
& \multicolumn{1}{c}{\multirow{4}{*}{\rotatebox[origin=c]{90}{$\infty$}}}
														& FBP   	& 1.0000			& 0.9296 			& 0.8383			& 0.7912			& 0.6844			& 0.5583			\\
& \multicolumn{1}{c}{}									& U-Net		& 0.9979			& 0.9937			& 0.9930			& 0.9904			& 0.9885			& 0.9703			\\
& \multicolumn{1}{c}{}									& W-Net		& 0.9976			& 0.9950			& 0.9951			& 0.9943			& 0.9925			& 0.9876			\\
& \multicolumn{1}{c}{}									& Proposed	& \textbf{0.9990}	& \textbf{0.9979}	& \textbf{0.9982}	& \textbf{0.9978}	& \textbf{0.9962}	& \textbf{0.9933}	\\ \cline{3-9} 

& \multicolumn{1}{c}{\multirow{4}{*}{\rotatebox[origin=c]{90}{$1.0 \times 10^7$}}}	
														& FBP   	& 0.9726			& 0.9058 			& 0.8119			& 0.7637			& 0.6488			& 0.5229			\\
& \multicolumn{1}{c}{}									& U-Net		& 0.9952			& 0.9922			& 0.9911			& 0.9880			& 0.9851			& 0.9663			\\
& \multicolumn{1}{c}{}									& W-Net		& 0.9952			& 0.9939			& 0.9935			& 0.9927			& 0.9897			& 0.9847			\\
& \multicolumn{1}{c}{}									& Proposed	& \textbf{0.9978}	& \textbf{0.9970}	& \textbf{0.9971}	& \textbf{0.9966}	& \textbf{0.9943}	& \textbf{0.9907}	\\ \cline{3-9} 

& \multicolumn{1}{c}{\multirow{4}{*}{\rotatebox[origin=c]{90}{$1.0 \times 10^6$}}}	
														& FBP   	& 0.7952			& 0.7578 			& 0.6560			& 0.6068			& 0.4915			& 0.3795			\\
& \multicolumn{1}{c}{}									& U-Net		& 0.9895			& 0.9874			& 0.9845			& 0.9800			& 0.9746			& 0.9517			\\
& \multicolumn{1}{c}{}									& W-Net		& 0.9906			& 0.9901			& 0.9886			& 0.9874			& 0.9821			& 0.9752			\\
& \multicolumn{1}{c}{}									& Proposed	& \textbf{0.9953}	& \textbf{0.9947}	& \textbf{0.9941}	& \textbf{0.9932}	& \textbf{0.9893}	& \textbf{0.9840}	\\ \cline{3-9} 

& \multicolumn{1}{c}{\multirow{4}{*}{\rotatebox[origin=c]{90}{$5.0 \times 10^5$}}}	
														& FBP   	& 0.6833			& 0.6702 			& 0.5688			& 0.5223			& 0.4191			& 0.3200			\\
& \multicolumn{1}{c}{}									& U-Net		& 0.9868			& 0.9845			& 0.9809			& 0.9756			& 0.9685			& 0.9435			\\
& \multicolumn{1}{c}{}									& W-Net		& 0.9886			& 0.9880			& 0.9860			& 0.9845			& 0.9782			& 0.9703			\\
& \multicolumn{1}{c}{}									& Proposed	& \textbf{0.9942}	& \textbf{0.9937}	& \textbf{0.9925}	& \textbf{0.9914}	& \textbf{0.9868}	& \textbf{0.9807}	\\ \cline{3-9} 

& \multicolumn{1}{c}{\multirow{4}{*}{\rotatebox[origin=c]{90}{$2.5 \times 10^5$}}}	
														& FBP   	& 0.5508			& 0.5714 			& 0.4750			& 0.4339			& 0.3477			& 0.2649			\\
& \multicolumn{1}{c}{}									& U-Net		& 0.9828			& 0.9803			& 0.9757			& 0.9694			& 0.9599			& 0.9316			\\
& \multicolumn{1}{c}{}									& W-Net		& 0.9859			& 0.9849			& 0.9826			& 0.9805			& 0.9732			& 0.9637			\\
& \multicolumn{1}{c}{}									& Proposed	& \textbf{0.9929}	& \textbf{0.9925}	& \textbf{0.9906}	& \textbf{0.9891}	& \textbf{0.9838}	& \textbf{0.9767}	\\ \cline{3-9} 

& \multicolumn{1}{c}{\multirow{4}{*}{\rotatebox[origin=c]{90}{$1.0 \times 10^5$}}}	
														& FBP   	& 0.3707			& 0.4442 			& 0.3631			& 0.3315			& 0.2684			& 0.2075			\\
& \multicolumn{1}{c}{}									& U-Net		& 0.9719			& 0.9698			& 0.9629			& 0.9548			& 0.9401			& 0.9045			\\
& \multicolumn{1}{c}{}									& W-Net		& 0.9798			& 0.9787			& 0.9763			& 0.9733			& 0.9638			& 0.9521			\\
& \multicolumn{1}{c}{}									& Proposed	& \textbf{0.9905}	& \textbf{0.9903}	& \textbf{0.9872}	& \textbf{0.9852}	& \textbf{0.9787}	& \textbf{0.9699}	\\ \cline{3-9} 
\hline \hline

\end{tabular}
\end{adjustbox}
\end{center}
\end{table*}

Fig. \ref{fig:result} shows the reconstructed results by FBP, U-Net, W-Net, and proposed network in terms of various low-dose ROI CTs. When the number of photons and the size of ROIs are too small, the U-Net does not remove the artifacts clearly, so the reconstructed images from U-Net remain distorted artifacts. Although the W-Net produces better results than the U-Net, the reconstructed images are so blurry in many areas that small structures and textures are not preserved. In contrast to the U-Net and the W-Net, which are image-domain CNNs, the proposed method preserves not only the small structures but also the textures of the images and shows the lowest NMSE values than the other methods. Despite both the W-Net and the proposed network use the two-times unrolled networks, the proposed network shows significant improvement over the W-Net. The important reason to improve the performance of the proposed network is that the proposed network uses a projection-domain CNN as the first network and satisfies the theory of deep convolutional framelets, whereas the W-Net does not meet the theory. 

\section{Discussion}
\label{sec:discussion}

\subsection{Network efficiency according to training domain}
\label{sec:net_eff}

\begin{figure*}[t!]
  \centering
  \includegraphics[width=0.95\textwidth]{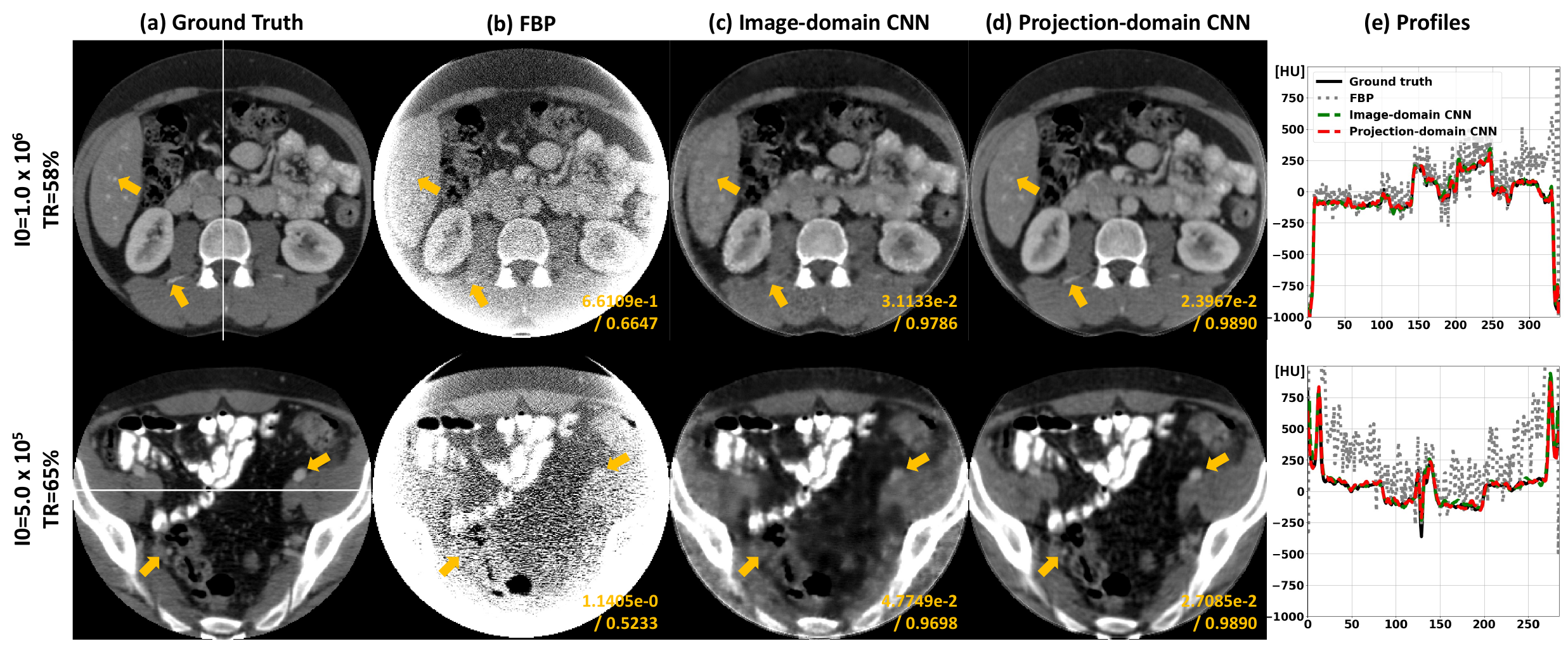}
  \caption{(a) Ground truth and reconstructed images by (b) FBP, (c) image-domain CNN and (d) projection-domain CNN. (e) Profiles along the white lines on the results. First and second rows show reconstructed images from $I0 = 1.0 \times 10^{6}$ with $58\%$ truncated detectors and $I0 = 5.0 \times 10^{5}$ with $65\%$ truncated detectors, respectively. The intensity range was set to $(-150, 400)$[HU]. NMSE / SSIM values are written at the corner. }
  \label{fig:result_eff}
\end{figure*}

\begin{table}[t!]
\caption{\bf\scriptsize Quantitative comparison between image- and projection-domain CNNs.}
\vspace*{-0.1cm}
\label{tbl:efficiency}
\begin{center}
\begin{adjustbox}{width=0.75\textwidth}
\begin{tabular}{cclcccccc}
\hline\hline
\multicolumn{3}{c}{\multirow{2}{*}{NMSE $(\times 10^{-2})$}}
															& \multicolumn{6}{c}{Ratios of truncated detectors} \\
\multicolumn{3}{c}{}										& $0\%$ 			& $44\%$ 			& $58\%$ 			& $65\%$			& $74\%$			& $79\%$			\\ \hline 
\multirow{12}{*}{\rotatebox[origin=c]{90}{The number of incident photons}}
& \multicolumn{1}{c}{\multirow{2}{*}{\rotatebox[origin=c]{0}{$\infty$}}}
												& I-domain	& 1.0780			& 1.8387			& 2.7758			& 3.5036			& 4.3792	        & 7.9992			\\
& \multicolumn{1}{c}{}							& P-domain	& \textbf{0.5346}	& \textbf{1.1330}	& \textbf{1.7984}	& \textbf{2.3472}	& \textbf{4.0934}	& \textbf{6.5316}	\\ \cline{3-9} 

& \multicolumn{1}{c}{\multirow{2}{*}{\rotatebox[origin=c]{0}{$1.0 \times 10^7$}}}	
												& I-domain	& 1.5831			& 2.0245			& 2.9153			& 3.6079			& 4.4841	        & 8.0629			\\
& \multicolumn{1}{c}{}							& P-domain	& \textbf{1.0127}	& \textbf{1.3645}	& \textbf{1.9474}	& \textbf{2.4774}	& \textbf{4.1746}	& \textbf{6.5956}	\\ \cline{3-9} 

& \multicolumn{1}{c}{\multirow{2}{*}{\rotatebox[origin=c]{0}{$1.0 \times 10^6$}}}	
												& I-domain	& 2.5743			& 2.7199			& 3.4837			& 4.0808			& 4.9567	        & 8.3683			\\
& \multicolumn{1}{c}{}							& P-domain	& \textbf{1.6421}	& \textbf{1.8032}	& \textbf{2.2945}	& \textbf{2.8109}	& \textbf{4.4102}	& \textbf{6.7657}	\\ \cline{3-9} 

& \multicolumn{1}{c}{\multirow{2}{*}{\rotatebox[origin=c]{0}{$5.0 \times 10^5$}}}	
												& I-domain	& 3.0332			& 3.1350			& 3.7999			& 4.3631			& 5.2635	        & 8.6015			\\
& \multicolumn{1}{c}{}							& P-domain	& \textbf{1.8966}	& \textbf{2.0086}	& \textbf{2.4715}	& \textbf{2.9798}	& \textbf{4.5435}   & \textbf{6.8480}	\\ \cline{3-9} 

& \multicolumn{1}{c}{\multirow{2}{*}{\rotatebox[origin=c]{0}{$2.5 \times 10^5$}}}	
												& I-domain	& 3.6433			& 3.7098			& 4.2361			& 4.7376			& 5.6669	        & 8.9502			\\
& \multicolumn{1}{c}{}							& P-domain	& \textbf{2.2099}	& \textbf{2.2664}	& \textbf{2.6960}	& \textbf{3.1917}	& \textbf{4.7167}   & \textbf{6.9653}	\\ \cline{3-9} 

& \multicolumn{1}{c}{\multirow{2}{*}{\rotatebox[origin=c]{0}{$1.0 \times 10^5$}}}	
												& I-domain	& 4.9174			& 4.8833			& 5.1807			& 5.5345			& 6.4666	        & 9.9097			\\
& \multicolumn{1}{c}{}							& P-domain	& \textbf{2.7906}	& \textbf{2.7455}	& \textbf{3.1177}	& \textbf{3.5908}	& \textbf{5.0470}	& \textbf{7.2115}	\\ \cline{3-9} 
\hline \hline

\multicolumn{3}{c}{\multirow{2}{*}{SSIM}}
															& \multicolumn{6}{c}{Ratios of truncated detectors} \\
\multicolumn{3}{c}{}										& $0\%$ 			& $44\%$ 			& $58\%$ 			& $65\%$			& $74\%$			& $79\%$			\\ \hline 
\multirow{12}{*}{\rotatebox[origin=c]{90}{The number of incident photons}}
& \multicolumn{1}{c}{\multirow{2}{*}{\rotatebox[origin=c]{0}{$\infty$}}}
												& I-domain	& 0.9973			& 0.9930			& 0.9895			& 0.9863			& 0.9856        	& 0.9697			\\
& \multicolumn{1}{c}{}							& P-domain	& \textbf{0.9994}	& \textbf{0.9967}	& \textbf{0.9961}	& \textbf{0.9941}	& \textbf{0.9889}   & \textbf{0.9784}	\\ \cline{3-9} 

& \multicolumn{1}{c}{\multirow{2}{*}{\rotatebox[origin=c]{0}{$1.0 \times 10^7$}}}	
												& I-domain	& 0.9946			& 0.9916			& 0.9877			& 0.9845			& 0.9822            & 0.9659			\\
& \multicolumn{1}{c}{}							& P-domain	& \textbf{0.9978}	& \textbf{0.9953}	& \textbf{0.9945}	& \textbf{0.9924}	& \textbf{0.9863}	& \textbf{0.9748}	\\ \cline{3-9} 

& \multicolumn{1}{c}{\multirow{2}{*}{\rotatebox[origin=c]{0}{$1.0 \times 10^6$}}}	
												& I-domain	& 0.9884			& 0.9866			& 0.9806			& 0.9774			& 0.9713        	& 0.9510			\\
& \multicolumn{1}{c}{}							& P-domain	& \textbf{0.9947}	& \textbf{0.9925}	& \textbf{0.9908}	& \textbf{0.9881}	& \textbf{0.9801}   & \textbf{0.9665}	\\ \cline{3-9} 

& \multicolumn{1}{c}{\multirow{2}{*}{\rotatebox[origin=c]{0}{$5.0 \times 10^5$}}}	
												& I-domain	& 0.9853			& 0.9833			& 0.9768			& 0.9734			& 0.9651			& 0.9424			\\
& \multicolumn{1}{c}{}							& P-domain	& \textbf{0.9932}	& \textbf{0.9911}	& \textbf{0.9889}	& \textbf{0.9859}	& \textbf{0.9767}	& \textbf{0.9621}	\\ \cline{3-9} 

& \multicolumn{1}{c}{\multirow{2}{*}{\rotatebox[origin=c]{0}{$2.5 \times 10^5$}}}	
												& I-domain	& 0.9805			& 0.9782			& 0.9714			& 0.9677			& 0.9566			& 0.9306			\\
& \multicolumn{1}{c}{}							& P-domain	& \textbf{0.9912}	& \textbf{0.9892}	& \textbf{0.9863}	& \textbf{0.9829}	& \textbf{0.9722}	& \textbf{0.9561}	\\ \cline{3-9} 

& \multicolumn{1}{c}{\multirow{2}{*}{\rotatebox[origin=c]{0}{$1.0 \times 10^5$}}}	
												& I-domain	& 0.9671			& 0.9654			& 0.9581			& 0.9546			& 0.9379			& 0.9045			\\
& \multicolumn{1}{c}{}							& P-domain	& \textbf{0.9863}	& \textbf{0.9850}	& \textbf{0.9807}	& \textbf{0.9763}	& \textbf{0.9623}	& \textbf{0.9423}	\\ \cline{3-9} 

\hline \hline 
\multicolumn{9}{c}{\scriptsize{\bf * I-domain and P-domain are defined as the first network for the W-Net and the proposed.}} \\
\hline \hline
\end{tabular}
\end{adjustbox}
\end{center}
\end{table}

In Sec. \ref{sec:result}, we showed that the proposed network outperforms the image-domain networks like U-Net and W-Net. The reason for its excellent performance is that the proposed method approaches it as a way to find decoupled solutions in projection domain even though the solution is coupled in image domain. To verify our claim, quantitative metrics are calculated for a single image-domain CNN $\mathscr{Q}_{img}$ (see Fig. \ref{fig:architectures}(a)) used as the first network of the trained W-Net and a single projection-domain CNN $\mathscr{Q}_{prj}$ (see Fig. \ref{fig:architectures}(b)) used as the first network of the trained proposed network, and the average NMSE and SSIM values are described in Table \ref{tbl:efficiency}. The projection-domain CNN $\mathscr{Q}_{prj}$ outperforms the image-domain CNN $\mathscr{Q}_{img}$, and Fig. \ref{fig:result_eff} shows reconstructed images for each network. Even though the reconstructed images from the image-domain CNN $\mathscr{Q}_{img}$ are not texture-preserved and suffer from severe blurring in many areas, the projection-domain CNN $\mathscr{Q}_{prj}$ removes the artifacts such as the image noise and the cupping artifact and preserves the underlying structure. 
For this result, because the image size is $512 \times 512$ but the projection data is $720 \times 1440$, we can doubts about an unfairness in the size of the training dataset between image domain and projection domain. However, the projection data can be transformed to the image data using a linear function such as the filtered backprojection (FBP) operation. From an information perspective, this means that the information of the projection data is similar to the information of the image data. 
Therefore, although many researchers use the image-domain CNN $\mathscr{Q}_{img}$ to solve CT problems, we argue that the projection-domain CNN $\mathscr{Q}_{prj}$, consisting of two bridge modules that estimate the projection noise inside measured region $\mathscr{T}$ and extrapolation map outside measured region $(1 - \mathscr{T})$, is more useful than the image-domain CNN $\mathscr{Q}_{img}$ because the projection domain provides a more efficient low-rank property than the image domain. In addition, the numerical results support our claim as shown in Table. \ref{tbl:efficiency} and Fig. \ref{fig:result_eff}.

\begin{figure*}[t!]
  \centering
  \includegraphics[width=0.5\textwidth]{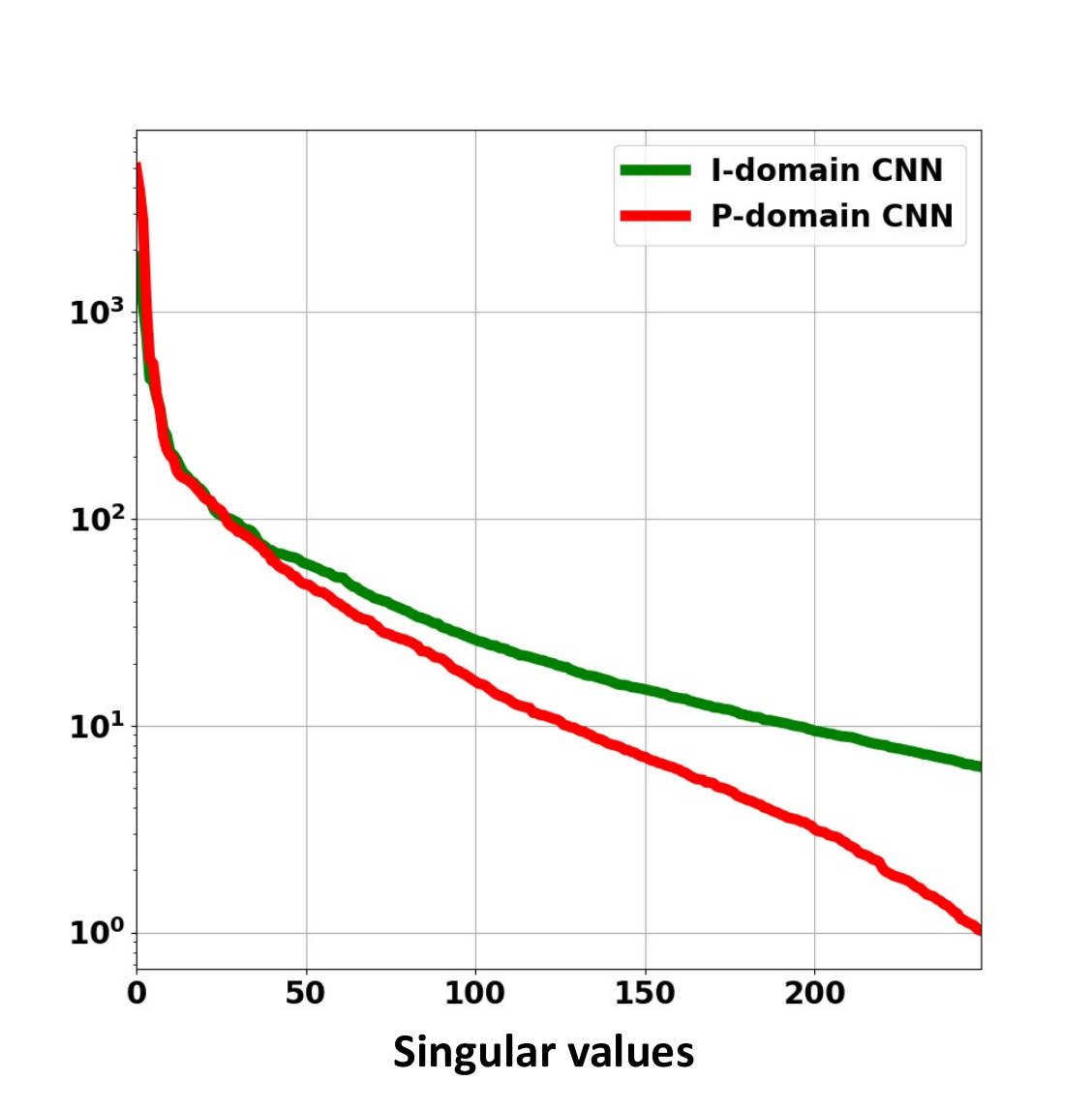}
  \caption{Singular value spectra of the Hankel structured matrix of the last feature maps of the backbone network according to the image-domain CNN (green) and the projection-domain CNN (red).}
  \label{fig:rank}
\end{figure*}

\subsection{Low rankness according to training domain}
\label{sec:low_rankness}
In the section \ref{sec:net_eff}, we confirmed that a projection-domain CNN shows better performance than an image-domain CNN. Here, to verify our claim that the projection-domain CNN can satisfy better low rank properties of the Hankel structured matrix than the image-domain CNN, we computed singular values of the Hankel structured matrix of the last feature maps of each backbone network, and the results were plotted in Fig. \ref{fig:rank}. As with the previous section \ref{sec:net_eff}, the projection-domain CNN performs better than the image-domain CNN in similar network architectures because the computed singular value spectra in the projected-domain CNN is lower than the image-domain CNN. Therefore, based on the Figs. \ref{fig:result_eff} and \ref{fig:rank} and Table. \ref{tbl:efficiency}, we verified that the projection-domain CNN is suitable for the low rankness and offers better performance than the image-domain CNN.

\subsection{Interior tomography vs. low-dose CT}

\begin{figure*}[t!]
  \centering
  \includegraphics[width=0.95\textwidth]{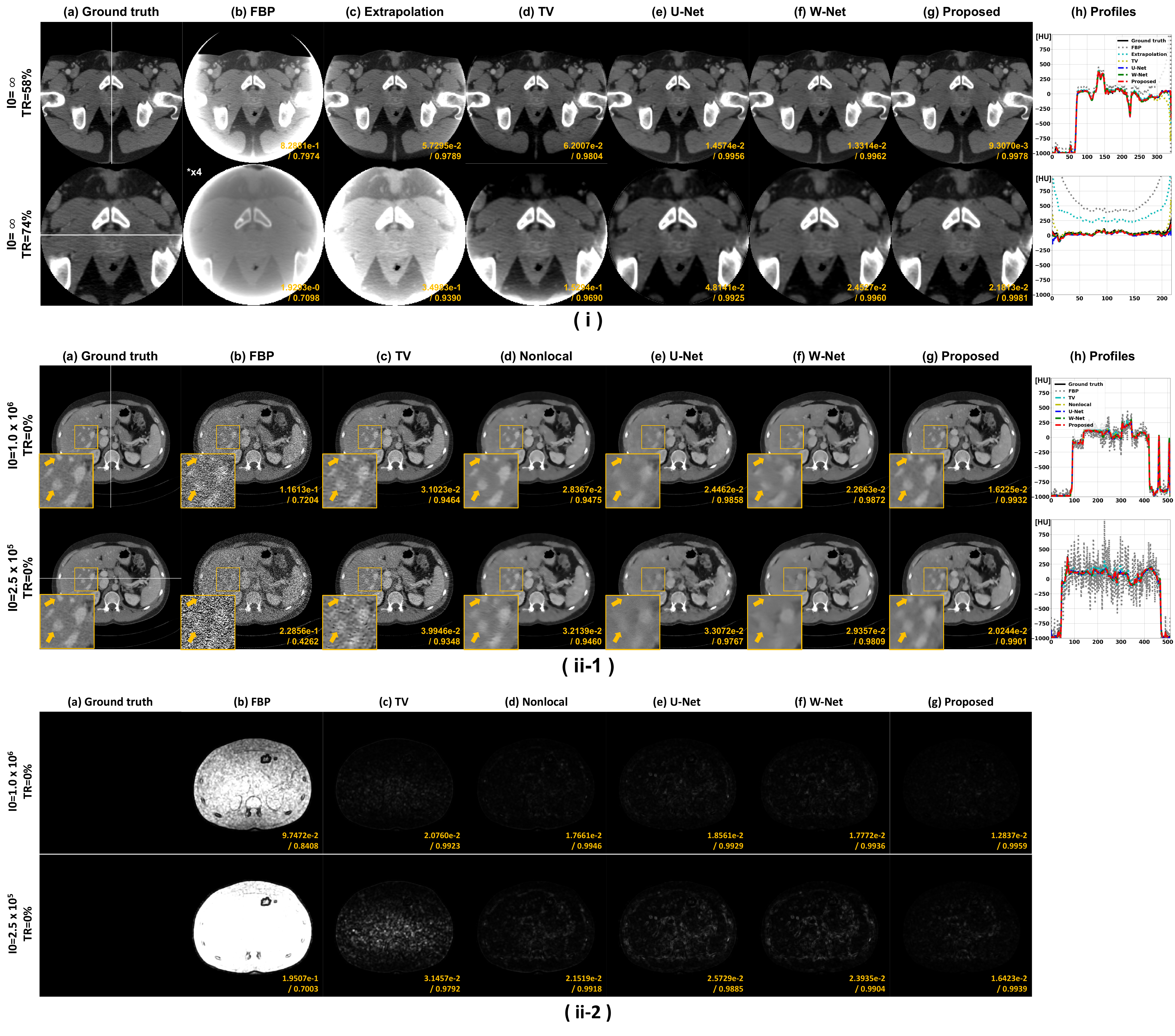}
  \caption{Reconstructed images from (i) interior tomography with $(58\%, 74\%)$ truncated ratios and (ii-1) low-dose CT with $I0 = (1.0 \times 10^6, 2.5 \times 10^5)$. (ii-2) SSIM maps calculated from in-body regions of (ii-1). The intensity range was set to $(-150, 400)$[HU]. $^{*}$x4 denotes that a window scale is magnified four times. NMSE / SSIM values are written at the corner.}
  \label{fig:single_task}
\end{figure*}

The proposed method was developed to solve a CT problem combining interior tomography and low-dose CT and showed superior performance than others as shown in Table. \ref{tbl:result} and Fig. \ref{fig:result}. Here, we can be wondering that how the performance differs when the problem is already isolated. Tables. \ref{tbl:roi_ct} and \ref{tbl:low_dose_ct} show PSNR values with respect to interior tomography without projection noise ($I0=\infty$) and low-dose CT without a detector truncation, respectively. Interestingly, the CT problem has already been decoupled into interior tomography and low-dose CT, but the proposed method is still superior to the other methods. Fig. \ref{fig:single_task} shows the reconstructed images from decoupled CT systems. 

For the interior tomography in Fig. \ref{fig:single_task}(i), the proposed network preserved details and textures of the images regardless of the truncated size. However, the U-Net and the W-Net showed high-quality reconstructed images when a ROI is large, but slightly blurry when the ROI is small. Extrapolation method, which is an existing method, was also performed to solve the interior tomography problem (see Fig. \ref{fig:single_task}(i)(c)). For large ROI, the extrapolation method moderately mitigated cupping artifacts, but for small ROI, it did not work. MBIR method with TV penalty showed more stable results than the extrapolation method, but did not exceed the DL performance. When the low-dose CT in Fig. \ref{fig:single_task}(ii), the U-Net and the W-Net removed image noise as well as small tissues, but the proposed method preserved the small tissues well, and only removed image noise. Specifically, MBIR method with TV penalty was performed to remove the image noise. Compared to the U-Net and the W-Net, the TV method preserves details better despite noise remaining in the reconstructed image as shown in Fig. \ref{fig:single_task}(ii)(c). Nonlocal prior method\cite{kim2017low}, which is the AAPM challenge winning algorithm, preserved details better than U-Net and W-Net and removed the image noises better than TV method. However, the NMSE and SSIM valuse do not seem to reflect the well-corrected image quality of the nonlocal prior method. To clarify the image quality evaluation when the low-dose measurement, the NMSE and SSIM values were re-calculated within body regions. Fig. \ref{fig:single_task}(ii-2) shows the SSIM map inside body regions, and the high-intensity represents the proportion of miss-matched structure with ground truth. Although the nonlocal prior method outperformed the TV, U-Net, and W-Net methods, the proposed method achieved best performance than the nonlocal prior method.


\subsection{Unrolled neural network architecture}
\label{sec:unrolled}


\begin{table*}[t!]
\caption{\bf\scriptsize Quantitative comparison with respect to interior tomography.}
\vspace*{-0.3cm}
\label{tbl:roi_ct}
\begin{center}
\begin{adjustbox}{width=0.8\textwidth}
\begin{tabular}{cclcccccc}
\hline\hline

\multicolumn{3}{c}{\multirow{2}{*}{PSNR [dB]}}		    & \multicolumn{6}{c}{Ratios of truncated detectors} \\
\multicolumn{3}{c}{}												& $0\%$ 			& $44\%$ 			& $58\%$ 			& $65\%$			& $74\%$			& $79\%$			\\ \hline 
\multirow{5}{*}{\rotatebox[origin=c]{90}{\scriptsize{Incident photons}}}
& \multicolumn{1}{c}{\multirow{5}{*}{\rotatebox[origin=c]{90}{$\infty$}}}
														& Extra.    & -			        & 36.1378			& 34.4970			& 27.3378			& 15.5640			& 8.7176			\\
&														& TV        & -			        & 38.4084			& 29.3277			& 29.0841			& 24.4248			& 18.7840			\\ \cline{3-9} 
& \multicolumn{1}{c}{}									& U-Net		& 50.5815			& 45.1681			& 42.0854			& 38.9915			& 35.0328			& 28.8829			\\
& \multicolumn{1}{c}{}									& W-Net		& 50.2332			& 46.0520			& 43.6709			& 41.9147			& 39.2676			& 34.5895			\\
& \multicolumn{1}{c}{}									& Proposed	& \textbf{53.3537}	& \textbf{50.4824}	& \textbf{47.6772}	& \textbf{46.3142}	& \textbf{41.0630}	& \textbf{37.2672}	\\

\hline \hline

\end{tabular}
\end{adjustbox}
\end{center}
\end{table*}

\begin{table*}[t!]
\caption{\bf\scriptsize Quantitative comparison with respect to low-dose CT.}
\vspace*{-0.3cm}
\label{tbl:low_dose_ct}
\begin{center}
\begin{adjustbox}{width=0.8\textwidth}
\begin{tabular}{cclcccccc}
\hline\hline

\multicolumn{3}{c}{\multirow{2}{*}{PSNR [dB]}}		    & \multicolumn{6}{c}{The number of incident photons} \\
\multicolumn{3}{c}{}												& $\infty$ 			& $1.0 \times 10^7$ & $1.0 \times 10^6$ & $5.0 \times 10^5$ & $2.5 \times 10^5$ & $1.0 \times 10^5$	\\ \hline 
\multirow{5}{*}{\rotatebox[origin=c]{90}{\scriptsize{Truncated rate}}}
& \multicolumn{1}{c}{\multirow{5}{*}{\rotatebox[origin=c]{90}{0\%}}}
														& TV		& -			        & 41.6881			& 39.9633			& 39.4789			& 37.4864			& 33.2091			\\
&														& Nonlocal  & -			        & 42.6110			& 41.3111			& 40.5736			& 39.5787			& 37.8072			\\ \cline{3-9} 
& \multicolumn{1}{c}{}									& U-Net		& 50.5818   		& 46.6985			& 42.5485			& 41.2001			& 39.6735			& 37.1377			\\
& \multicolumn{1}{c}{}									& W-Net		& 50.2332			& 46.8628			& 43.1995			& 41.9968			& 40.6882			& 38.6591			\\
& \multicolumn{1}{c}{}									& Proposed	& \textbf{53.3537}	& \textbf{50.0518}	& \textbf{46.5811}	& \textbf{45.5246}	& \textbf{44.4409}	& \textbf{42.8878}	\\  

\hline \hline

\end{tabular}
\end{adjustbox}
\end{center}
\end{table*}

The single CNNs in Figs. \ref{fig:architectures}(a, b) are recognized as one-time unrolled CNNs, even though their training domains are different. From the previous point of view, the W-Net in Fig. \ref{fig:architectures}(c) and the proposed network in Fig. \ref{fig:architectures}(d) are the two-times unrolled CNNs. However, the W-Net is stretched only in the image domain, but the proposed network is unfolded sequentially in the projection domain and the image domain. Comparing Table. \ref{tbl:result} and Table. \ref{tbl:efficiency}, the two-times unrolled CNNs outperforms the one-time unrolled CNNs. Since the two-times unrolled CNNs have twice the parameter size than the one-times unrolled CNNs, it can be taken for granted. However, the U-Net has two-times the parameters of the W-Net, but the W-Net shows better performance than the U-Net. In a comparative study of U-Net and W-Net, we found that a sequence of several DLs with low expressibility shows better performance than a single DL with high expressibility.
In addition, both the W-Net and the proposed network are the two-times unrolled CNN, but as confirmed in Section \ref{sec:net_eff}, the proposed method initially unfolded in projection domain is more efficient than the W-Net. 

\section{Conclusion}
\label{sec:conclusion} 
In this paper, we proposed the novel and efficient low-dose interior CT reconstruction. Due to the coupled artifacts, we have shown that image-domain CNNs do not satisfy the low rankness, which can degrade the performance. Thus, instead of using the image-domain CNN, we decoupled inside and outside measured regions in the projection domain to efficiently estimate separated solutions. To address the decoupling, we proposed the novel end-to-end deep learning method consisting of projection- and image-domain CNNs referred to as the dual-domain CNNs. Specifically, the projection-domain CNN solves two major problems; (1) noise from inside the measured region and (2) extrapolation from outside the measured region. Then, the image-domain CNN further improved the image quality. Our results demonstrated that the proposed method outperformed the model-based reconstruction methods and the conventional image-domain CNNs.

\section*{References}



\bibliographystyle{unsrt}
\bibliography{ref.bib}

\end{document}